\documentclass{lsemulateapj}
\pdfoutput=1
\usepackage{pdflscape}
\usepackage{graphicx,natbib}
\usepackage{amsmath}
\usepackage{hyperref}

\newcommand{\ace}{$\alpha_{\rm CE}$}
\newcommand{\frlof}{$f_{\rm RLOF}$}
\newcommand{\bse}{\texttt{BSE}}

\begin{document}
\title{Population Synthesis of Hot Subdwarfs: A Parameter Study}
\author{Drew Clausen}
\author{Richard A Wade}
\affil{Department of Astronomy and Astrophysics, 525 Davey Lab, The Pennsylvania State University, University Park, PA 16802, USA}
\author{Ravi Kumar Kopparapu}
\affil{Virtual Planetary Laboratory, Department of Geosciences, 443 Deike Building, The Pennsylvania State University, University Park, PA 16802, USA}
\author{Richard O'Shaughnessy}
\affil{Center for Gravitation and Cosmology, 462 Physics Building, University of Wisconsin-Milwaukee, Milwaukee, WI 53211, USA}
\email{dclausen@astro.psu.edu, wade@astro.psu.edu, ravi@gravity.psu.edu, oshaughn@gravity.phys.uwm.edu}
\shorttitle{Population Synthesis of sdBs}
\shortauthors{Clausen et al.}
\slugcomment{Accepted for publication in \apj}

\begin{abstract}
Binaries that contain a hot subdwarf (sdB) star and a main sequence companion may have interacted in the past.  This binary population has historically helped determine our understanding of binary stellar evolution.  We have computed a grid of binary population synthesis models using different assumptions about the minimum core mass for helium ignition, the envelope binding energy, the common envelope ejection efficiency, the amount of mass and angular momentum lost during stable mass transfer, and the criteria for stable mass transfer on the red giant branch and in the Hertzsprung gap.  These parameters separately and together can significantly change the entire predicted population of sdBs.  Nonetheless, several different parameter sets can reproduce the observed subpopulation of sdB + white dwarf and sdB + M dwarf binaries, which has been used to constrain these parameters in previous studies. The period distribution of sdB + early F dwarf binaries offers a better test of different mass transfer scenarios for stars that fill their Roche lobes on the red giant branch.                     
\end{abstract}
\section{Introduction}
Binary population synthesis (BPS) models parameterize several physical processes that can drastically alter many properties of the stellar populations that they predict \citep{Eggleton:2006}.  While some of these parameterizations are used to reduce the computational burden of evolving $\ga 10^{6}$ binaries, many stem from our incomplete understanding of single and binary star evolution.  These phenomenological parameterizations are tuned to reproduce well-known binary populations \citep[e.g.][]{Nelemans:2000}.  Binaries containing a subdwarf B (sdB) star constitute one often-used population.  These objects, also called extreme horizontal branch (EHB) stars, are thought to be core helium burning stars with very thin hydrogen envelopes, $M_{\rm env} \la 0.02~M_{\sun}$, \citep[for a recent review, see][]{Heber:2009}.  Such  objects can be formed in
interacting binaries, by stripping  the hydrogen envelope from a star as it ascends the red giant branch \citep{Mengel:1976}.    Observations support a binary origin for a significant fraction of sdBs. Not only are most known sdBs in binaries with tight periods $P\la10$ days \citep{Maxted:2001}, but their semimajor axis distribution is inconsistent with the primordial distribution proposed by \citet{Duquennoy:1991} \citep{Heber:2002}.  Being ubiquitous,  bright, and produced by interacting binaries, hot subdwarfs  are an ideal population to test binary-evolution scenarios  \citep{Green:2001,Han:2003,Nelemans:2010}. 

Several previous studies have investigated sdB formation with a theoretical approach \citep[e.g.][]{Tutukov:1990,Yungelson:2005,Nelemans:2010}. \citet{Han:2002} computed detailed binary evolution models for five sdB formation channels: the first and second common-envelope (CE) ejection channel, the first and second stable Roche lobe overflow (RLOF) channel, and the double helium white dwarf (WD) merger channel.  The results of these calculations were then applied in a BPS study presented in \citet[][hereafter H03]{Han:2003}.  The BPS model was able to reproduce certain properties of the observed population, including the binary fraction and the masses and periods of the sdB + WD binaries reported in \citet{Maxted:2001} and  \citet{Morales-Rueda:2003}.  Despite the success of many aspects of this model, some properties of the modeled population are not consistent with a recent sdB radial velocity study discussed in \citet{Copperwheat:2011}. These authors found that the binary fraction among sdBs is 50\%, but it is the absence of short period ($P<10$ days)  sdB + G or K type dwarf binaries from their sample that is most incompatible with the predictions of H03.  For the sdBs that showed evidence of a G or K type companion in the \citet{Copperwheat:2011} sample, no radial velocity variations were detected on short (few day) time scales, which implies that these systems have long periods.  They concluded that the presence of a G or K type companion is indicative of a long period binary,  which is consistent with the sdB radial velocity survey carried out by \citet{Green:2001} who inferred that $P \sim 1 - 3$ years for such systems.  Only one short period sdB + G or K dwarf system has been reported, \citet{Moni-Bidin:2010} identified such a binary in the globular cluster NGC6752 with $P \la 5$ days.  This is in stark contrast to the large population of sdB + G or K type systems with $P < 10$ days predicted by H03, see, e.g., their Figure 15.  

To explore the discrepancy between the models and observations and to test how the assumptions made in binary stellar evolution models impact the sdB population, we have carried out our own BPS study of sdB formation.  In \autoref{sec:bps} we describe our BPS calculation.  The results of our calculation are discussed in \autoref{sec:results}.  We show the initial binary distributions that lead to sdBs with hydrogen-burning companions; we show the joint distribution of orbital periods and companion effective temperatures at the present epoch; and we discuss which combinations of BPS parameters are compatible with observations of sdB + WD and sdB + M dwarf populations.  We describe how sdB + early F binaries can better constrain the models in \autoref{sec:constraints}. In \autoref{sec:comp} we compare our models with previous results.  Throughout, we distinguish between true helium-burning sdB stars and post-RGB stars with similar effective temperatures and surface gravities.  

\section{Binary Population Synthesis}
\label{sec:bps}
In this work, we focus on binaries in which the primary, the initially more massive star, has evolved off of the main sequence (MS), expanded to fill its Roche lobe, and lost its envelope to become an sdB.  The envelope can be lost either through stable RLOF mass transfer to the secondary or through CE ejection.  In these systems, the secondaries are typically MS stars, but some will have evolved as far as the red giant branch (RGB).  For simplicity we will often refer to all binaries in which the sdB has a hydrogen burning companion (both core-burning MS stars and shell-burning RGB stars) as sdB + MS binaries.   Although we were interested in the population of sdBs with hydrogen burning companions,  we also tracked sdB + WD binaries, in which the secondary has evolved to become the sdB, because nearly all of the observational constraints come from this subpopulation.        

\subsection{Binary Evolution Code}
\label{bse}
We used the binary evolution code \texttt{BINARY STAR EVOLUTION (BSE)} described in \citet{Hurley:2002}.  Below we describe several aspects of \bse~that are relevant to our study, but we refer the interested reader to \citet{Hurley:2002} for a complete description of the code.  In our models, we varied several parameters that govern a system's evolution to explore how each of these components impacted the resultant population of sdB + MS  binaries (see \autoref{table}).  First, we varied the minimum core mass at which helium ignites.  By default in \bse,  stars that undergo a helium flash (effectively stars with $M_{\rm ZAMS} <  1.995~ M_{\sun}$ at solar metallicity) do not ignite helium if their envelopes are lost before they reach the tip of the RGB (TRGB).  However, \citet{DCruz:1996} showed that it is possible for these stars to ignite helium if the envelope is lost within $\sim 0.4 $ bolometric magnitudes of the TRGB.  Stellar evolution models described in \citet{Han:2002} suggest that if the mass loss is rapid, as is the case for CE ejection, stars can ignite helium if the core has reached 95\% of mass it would have at the TRGB ($m_{\rm c}({\rm TRGB})$) before the envelope is removed.  We modified \bse~ to allow stars to ignite helium if the core has reached a mass 
\begin{equation}
m_{\rm c}=f_{\rm He}\;m_{\rm c}({\rm TRGB})
\end{equation}
before the star's envelope is lost, according to the result of \citet{Han:2002}.  As seen in \autoref{table}, we constructed synthetic populations with $f_{\rm He}=0.95$ that allowed helium ignition in stars that lose their envelopes shy of the TRGB, through either CE ejection or stable RLOF, and models with $f_{\rm He}=1$ that required the stars to reach the TRGB before helium ignition. 

Second, when the donor star fills its Roche lobe, the mass ratio $q = M_{\rm d}/M_{\rm a}$ determines whether or not mass transfer is dynamically stable.  Here $M_{\rm d}$ is the mass of the donor star and $M_{\rm a}$ is the mass of the accretor.  Stability of mass transfer depends on the structure of the donor,  the mass ratio of the binary, and other factors like the amount of mass and angular momentum lost to infinity \citep{Hjellming:1987,Soberman:1997,Webbink:2006,Ge:2010}.  When $q$ is smaller than a critical value $q_{\rm crit}$ the mass transfer is stable, otherwise the system will undergo CE evolution.  The functional form of the critical mass ratio versus all stellar and binary parameters is poorly known. To gauge how different assumptions about $q_{\rm crit}$ influenced the population of sdB + MS binaries, we considered many different values for this parameter.  For systems undergoing RLOF while the donor is in the Hertzsprung gap (HG), we set $q_{\rm crit}$ to either 3.2 or 4 \citep[][ respectively]{Han:2003,Hurley:2002}.  If the system undergoes RLOF while the primary is on the RGB, \bse's default critical mass ratio is given by
\begin{equation}
\label{eqn:qcrit}
q_{\rm crit} = 0.362 + \frac{1}{3(1-\frac{m_{\rm cd}}{M_{\rm d}})}
\end{equation}
where $m_{\rm cd}$ is the donor's core mass and $M_{\rm d}$ is the total mass of the donor star.  We have used the value of $q_{\rm crit}$ given by \autoref{eqn:qcrit} in some of our BPS models.  Under this assumption, $q_{\rm crit} < 1$  in most cases and mass transfer is unstable. \autoref{eqn:qcrit} was derived from models of conservative mass transfer between two condensed polytropes by \citet{Hjellming:1987}, however \citet{Han:2002} described a set of detailed binary stellar evolution calculations for completely non-conservative mass transfer in which the mass lost from the system carries away the specific angular momentum of the accretor.  Under these conditions, mass transfer was stable for $q \la 1.2$.  H03 argued that enhanced wind mass loss would have the effect of increasing $q_{\rm crit}$.  Following this suggestion, we also computed models with $q_{\rm crit} = 1.5$ on the RGB.

Third, in addition to changing the criteria that determine whether or not mass transfer will be dynamically stable, we varied parameters that govern each mass loss scenario.  During stable RLOF mass transfer, some of the mass lost by the donor might not be accreted by the companion, but instead will be lost from the system.  We defined \frlof~as the fraction of mass lost by the primary that is accreted by the secondary during stable RLOF mass transfer, e.g.,  \frlof ~= 1 corresponds to conservative mass transfer.  By default in \bse, the amount of material accreted by a MS or HG secondary is limited by its thermal timescale and \frlof~is computed with
\begin{equation}
\label{eqn:frlof}
f_{\rm RLOF} = \min\left(1.0, 10\,  \frac{M_{\rm a}}{\dot{M_{\rm d}}\tau_{\rm KHa}}\right)
\end{equation}
where $M_{\rm a}$ is the mass of the accretor, $\dot{M_{\rm d}}$ is the mass loss rate of the donor, and $\tau_{\rm KHa}$ is the Kelvin-Helmholtz timescale of the accretor.  Setting $f_{\rm RLOF}$ with \autoref{eqn:frlof} allows the accretor to accept all of the material lost by the donor for sufficiently low $\dot{M_{\rm d}}$, but caps the accretion rate at a value that prevents the accretor's radius from expanding by more than a factor of $\sim 1.5$ for larger $\dot{M_{\rm d}}$ \citep{Tout:1997,Pols:1994}.  We also computed models with \frlof ~= 0.5, again, following H03.  

Fourth, the mass lost by the primary that is not accreted by the secondary leaves the system and it must also carry away angular momentum.  This material can take angular momentum from the orbit, the donor, or the accretor.  In the first case, denoted by  $\gamma = 1$ in the text and \autoref{table}, the material takes away a specific angular momentum proportional to the orbital angular momentum of the system, $a^{2}\Omega_{\rm Orb}$ where $\Omega_{\rm Orb}$ is the orbital angular frequency and $a$ is the semimajor axis.  In the second case, denoted by $\gamma = -1$, the material takes away a specific angular momentum proportional to that of the donor, $a_{\rm d}^{2}\Omega_{\rm Orb}$.  Cases in which the material carries away a specific angular momentum proportional to that of the accretor,  $a_{\rm a}^{2}\Omega_{\rm Orb}$, are denoted by $\gamma = -2$.  Note that $q_{\rm crit}$ and the parameters that describe non-conservative mass transfer, $f_{\rm RLOF}$ and $\gamma$, are coupled for fixed assumptions about stellar structure, but in our parameter study we treat them as independent inputs.     

Finally, we also considered several different values for the parameters that govern CE evolution.  During a CE phase, the mass donor's envelope drastically expands and engulfs  the accretor.  While this process is not well understood, it is believed that the core of the donor and the entire accretor spiral towards one another within the CE as orbital energy goes into heating and unbinding the CE.  This process drastically reduces the binary's orbital period, and often results in a merger.   In \bse~this process is controlled by two parameters.  The structure parameter $\lambda$ determines the envelope binding energy and \ace~determines how efficiently the orbital energy is transferred to the envelope.  We considered two cases for the envelope binding energy.  One was the \bse~default, which is to determine the value of $\lambda$ from the star's structure and evolutionary state using fitting formulae derived from single star evolution models.  With the value of $\lambda$ the binding energy (BE) is then calculated as $E_{\rm bind}= GM_{1}(M_{1}-m_{\rm c1})/(\lambda R_{1})$, where $R_{1}$ is the radius of the mass donor.   Alternatively, we used the analytic expressions for the envelope binding energy given in \citet[][hereafter  LVK11]{Loveridge:2011}.  These authors computed the envelope binding energy from a grid of detailed stellar evolution models and fit it as a function of metallically, mass, radius, and evolutionary phase.  The fits we used included terms for the internal energy of the gas, the radiation energy, and the gravitational energy, but not the recombination energy.  As the fraction of the orbital energy available to eject the common envelope, on physical grounds $\alpha_{\rm CE}<1$.   Treating the CE parameter \ace~as a purely phenomenological factor, we considered  $\alpha_{\rm CE}=$ 0.75, 1.5, and 3.  

\subsection{Initial and Present Day Population}
\label{sec:inipop}
The initial population of main sequence binary systems, some of which eventually have an sdB companion, was chosen as follows. To ensure that the distribution of sdB binaries is well sampled, we selected $3.5 \times 10^{6}$ initial binary systems, with the mass of both the primary ($M_{10}$, more massive) and secondary ($M_{20}$, less massive) in the  
range $0.1~M_\odot - 10~M_\odot$. The primary's mass distribution was taken from the initial mass function (IMF) of \citet{Kroupa:2003}, as given in \citet{Hurley:2002}
\begin{equation}
\label{eqn:massfunction}
\xi\left( m \right) = \begin{cases}
0, &m \le m_{0} \\
b_\mathrm{1} m^{-1.3}, & m_\mathrm{0} < m \le 0.5 \\
b_\mathrm{2} m^{-2.2}, & 0.5 < m \le 1.0 \\
b_\mathrm{2} m^{-2.7},& 1.0 < m < \infty
\end{cases}
\end{equation}
where $\xi(m) dm$ is the probability that a star has a mass between $m$ and $m+ dm$ and we have used $m_{0} = 0.1~M_{\sun}$. The initial mass ratio ($1/q_{0}$) distribution was taken to be uniform between $0$ and $1$, so the secondary's mass was correlated with the primary by $M_{2} =  M_{1}/q_{0}$. The initial mass ratio distribution is not well constrained by observations, but has a significant impact on the results.  For example, H03 considers both a constant mass ratio distribution case and a case in which each member of the binary is chosen form the IMF independently.  In the uncorrelated mass distribution case, the number of systems formed through CE ejection increased drastically.  However, an uncorrelated mass ratio distribution is disfavored by observations \citep[e.g.][]{Duquennoy:1991}.  We did not systematically explore the mass ratio distribution, and chose to use the constant mass ratio distribution in all of our runs to enable comparisons with H03's best fit model, which also used a uniform distribution of $1/q_{0}$.  We note, however, it is possible to rescale any simulation to an arbitrary mass ratio distribution. The distribution of initial orbital separation ($a$) was taken to be uniform in $\ln (a)$ between $3 ~R_\odot$ and $10^{4}~R_\odot$ \citep[][and references therein]{Hurley:2002}.  All orbits were assumed to be circular and all stars were assumed to have solar metallicity.

The same initial population of main sequence binaries was given as an input to the \bse~code for each run. The code uses an adaptive integrator to sample the evolutionary tracks of each binary.  In order to obtain the present day distribution of sdB binary systems produced during the evolution of these initial main sequence binaries, we resampled the output to a fixed sampling rate, linearly interpolating the characteristics of each individual star (gravity, effective temperature, luminosity etc.) so that we have a datapoint for every 10 Myr, starting from its birth, until either the maximum evolution time (15 Gyr in our case) or until one or both of the stars had no remnants. We then selected systems in which one or both the stars are sdBs.

Since we cannot observe whether a star is burning helium in its core, we chose to identify sdBs by their positions in $(T_{\rm eff}, \log g)$ space.  For this work we defined the ``hot subdwarf box'' to be the region where $20000 < T_{\rm eff} < 45000$ K and $5.0 < \log g < 6.6$ (cgs). Once a star enters the ``hot subdwarf box'' we record the resampled evolutionary history, while the system remains in an sdB phase. The collective evolutionary record of all such systems that are in the sdB phase then mimics a sample of present day sdB binaries that are in various stages of evolution. Using this method, the lifetimes of systems in an sdB phase are appropriately reflected in the present day population. If an sdB binary has a short lifetime, it remains in the sdb ``box'' for fewer resampled timesteps and hence such a system contributes less to the total number of current systems.  However, we do not specify a star formation rate and defer discussion of the birthrates or space densities of the binaries produced in our BPS models to future work.

Furthermore, we only consider sdBs with companions.  There are many proposed formation channels for single sdBs, including the merger of two He white dwarfs (WDs)
\citep{Webbink:1984,Iben:1984,Iben:1986}, enhanced RGB mass loss \citep{DCruz:1996},  ejection of the H envelope by a sub-stellar companion \citep{Soker:1998}, centrifugally enhanced mass loss triggered by common envelope (CE) mergers \citep{Politano:2008}, and the merger of an M dwarf with a He WD \citep{Clausen:2011}.  However, the algorithms in \bse~either do not include or do not self-consistently model these mergers and the evolution of the remnant.  We therefore draw no conclusions about the fraction of sdBs that are in binaries.      
\begin{figure}
	\centering
	\includegraphics[width=0.45\textwidth]{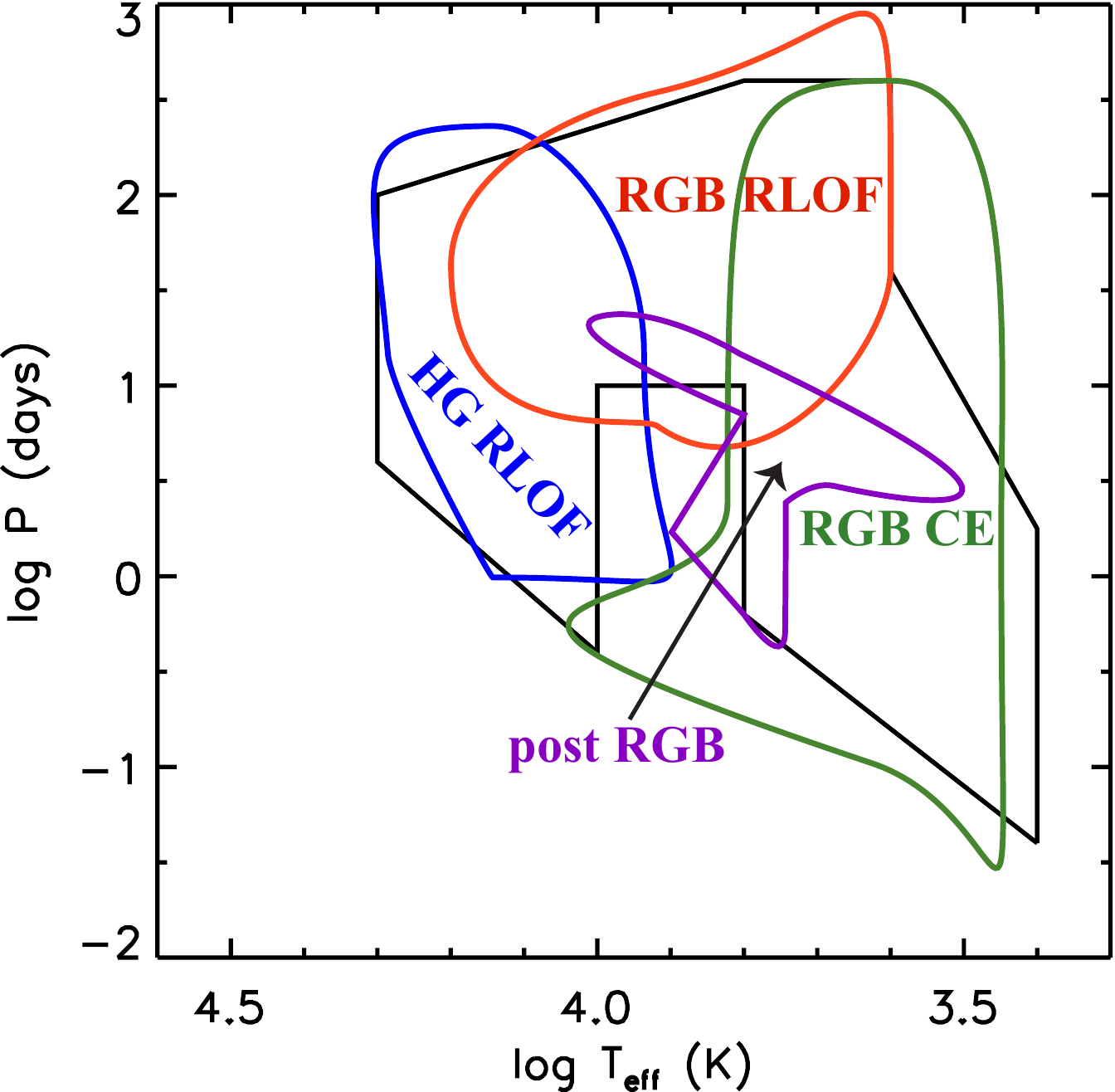}
	\caption{Cartoon of the regions of $\log P -\log T_{\rm eff}$ plane occupied by sdB binaries produced by different evolutionary channels.  Here $\log T_{\rm eff}$ refers to the temperature of the MS companion.  Note that these regions are the composite of all of our runs.  With a particular set of parameters, only a subset of these regions might appear and only a subregion of each region may be populated.\label{fig:han_toon}}
\end{figure}
	
We note that the mass transfer and stellar evolution algorithms used in \bse~do not produce true sdBs, instead they produce naked helium stars that have had the entire hydrogen envelope removed.  These naked helium stars all fall along the He main sequence, which runs through the high $\log g$, high $T_{\rm eff}$ corner of our sdB box.  Using the models of \citet{Caloi:1972}, we have confirmed that these naked helium stars would still be in the ``hot subdwarf box'' if they had hydrogen envelopes of up to $ 0.02~M_{\sun}$.  In the most extreme case, adding a $0.02~M_{\sun}$ envelope to the least massive star in our models ($M = 0.32~M_{\sun}$), $\log g$ would decrease by 1.2 dex, $\log T_{\rm eff}$ would decrease by 0.2 dex, and the star would still be in the ``hot subdwarf box.'' Although \bse~does not produce sdBs with hydrogen envelopes, we have only selected systems that would still be in the proper $\log g$ and $T_{\rm eff}$ range if they had a thin hydrogen envelope.  Because of this limitation of \bse, we do not discuss the $T_{\rm eff}$ or $\log g$ distributions of the sdB stars themselves.  Nor without further assumptions can we discuss the radiative properties (magnitudes, colors) of the sdBs or the binaries in which they are found.  The sdB masses (with small errors from neglecting the H envelope mass) remain valid for discussion, however.              

\section{Results}
\label{sec:results}
The sdB binary populations produced in our BPS models are summarized in \autoref{table}.  Column  (1) gives the identification number for the run and the next seven columns list the parameters and parameterizations used in the run.  Column (9) gives the number of initial binaries that produce an object that lies in the sdB box and column (10) lists the number of resampled, present day binary systems containing such an object.  Column (16) gives the percentage of the resampled systems that consist of a core helium burning sdB with a hydrogen burning companion and the preceding five columns give the percentage of systems formed by each formation channel.   Column (17) lists the percentage of core helium burning sdB binaries with WD companions.  Column (18) lists the percentage of present day systems that do not contain a true helium-burning sdB but instead an post-RGB star that is presently in the sdB box on its way to becoming a WD.  The model suitability parameter $\rho$ (see \autoref{section:gen}) is listed in column (19).
\begin{figure}
	\centering
	\includegraphics[width=0.45\textwidth]{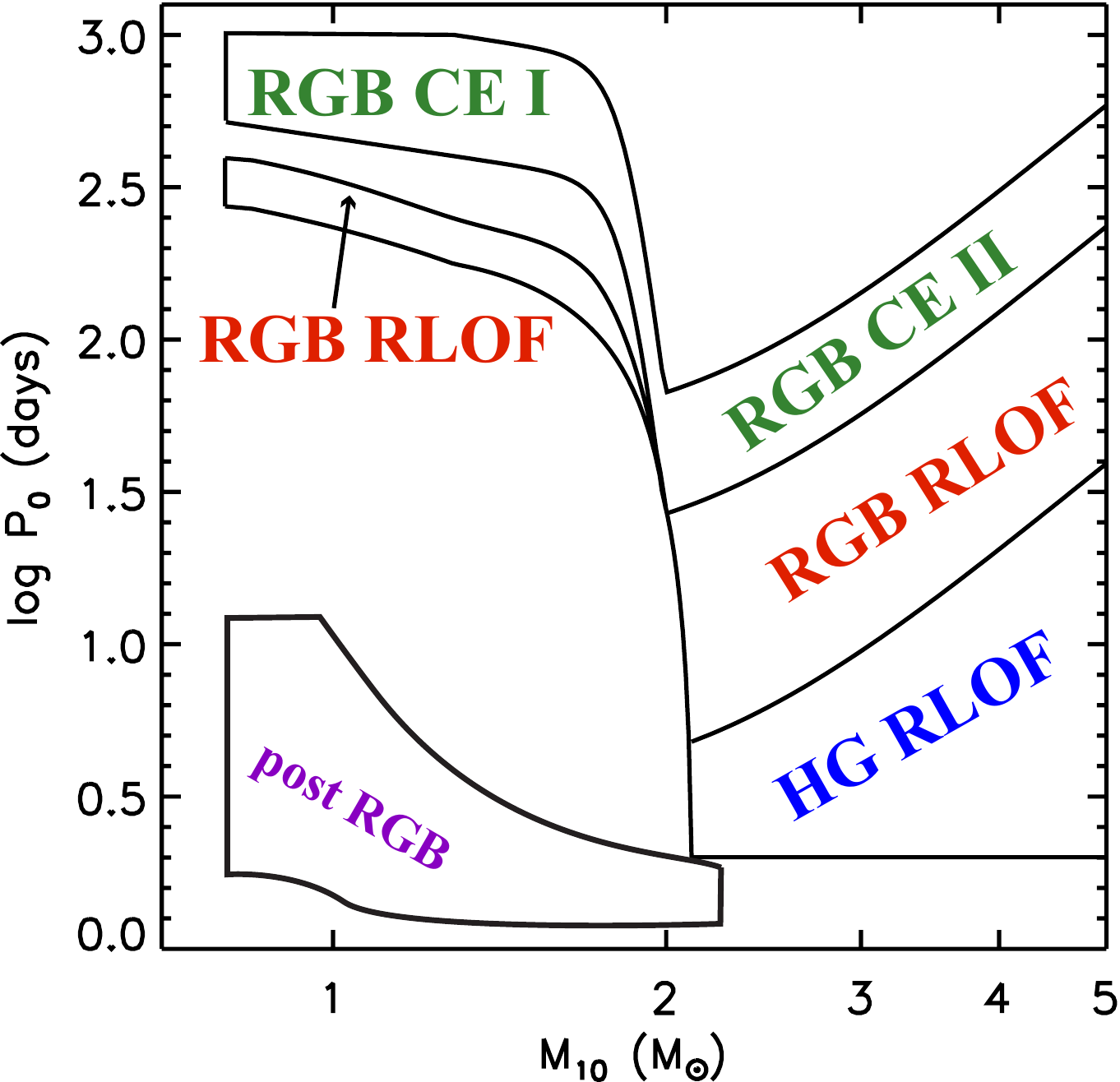}
	\caption{Cartoon of regions of the $\log P_{0}-M_{10}$ plane that produce helium-burning sdB + MS binaries through four evolutionary channels and the region that produces post-RGB stars that sit in the sdB ``box''.\label{fig:ic_toon}}
\end{figure}

Our BPS models show that the population of sdB + hydrogen burning companion binaries is highly sensitive to the value of several parameters necessary to model binary evolution.  Varying the values \ace, $\gamma$, $q_{\rm crit}$, $f_{\rm RLOF}$, $f_{\rm He}$, and the envelope binding energy, within reasonable ranges for each, resulted in substantial changes to the distributions of both the orbital periods and companion masses for sdB + MS binaries. Furthermore, our models suggest that these parameters are poorly constrained by the current sample of sdB + WD and sdB + M dwarf binaries with measured orbital periods because many of our models produced a population of such systems that was consistent with the observed sample. Assumptions about the IMF, the distribution of initial orbital periods and eccentricities, and the distribution of the initial mass ratios also affect the the populations and our conclusions, but were not varied among the different models presented.  

\subsection{General Description}  
\label{section:gen}
Before we discuss and compare the populations produced by individual runs in detail, we present a general description of the results and outline how we will present, examine, and evaluate each run.  To display the observable properties of these predicted populations, we will plot each binary's orbital period against the effective temperature ($T_{\rm eff}$) of the MS companion in $\log P - \log T_{\rm eff}$ diagrams like \autoref{fig:han_toon}.  The black lines outline the region occupied by the sdB + MS binaries predicted by the best-fitting BPS model chosen in H03 (see their Figure 15b).  Also shown are sketches of the regions occupied by sdB + MS binaries formed through different formation channels if we \emph{combine all} of our BPS models.  The regions shown are schematic; the exact boundaries of these regions and whether or not they are populated for a particular run depend on the parameters used in that run.  

Systems that form as the result of stable mass transfer beginning while the primary is in the HG (HG RLOF) tend to have the most massive companions because up to $3.5~M_{\sun}$ of material is transferred to the companion.  However, the period distribution of these systems is sensitive to assumptions about the amount $(f_{\rm RLOF})$ and specific angular momentum ($\gamma$) of material lost to during mass transfer.  

If we set $q_{\rm crit} = 1.5$ on the RGB, then sdB + MS binaries can form through stable mass transfer that begins on the RGB (RGB RLOF).  In these binaries, the MS companions span a large range in temperature.  This is due, in part, to evolution of the companion towards the RGB while the primary is an sdB.  The periods of these systems are typically longer than 5 days.  Some binaries survive CE evolution that begins on the RGB (RGB CE) to form sdB + MS binaries.  In the vast majority of these systems, the companion is of type G or later.  Note that the H03 models presented predict systems with companions to a lower $T_{\rm eff}$ than our models do.  This is a consequence of our choosing $m_{0} = 0.1$.  If we instead take $m_{0} = 0.08$, then the $\log~T_{\rm eff}$ range extends to 3.33.  Since we do not draw $M_{2}$ from the IMF, but determine it as $M_{2} = q_{0}M_{1}$, there are few systems with companion masses this low. 

We selected sdBs from the BPS runs based on their surface gravity and effective temperature, much as an observer would.  By this process, we also selected many objects that were not core helium burning sdBs, but the cores of post-RGB stars cutting across the ``hot subdwarf box.''  These objects are less massive than helium burning sdBs and have an average mass of $0.2~M_{\sun}$.  Individually, these systems spend little time in the sdB ``box,'' however they are quite numerous.  Interestingly, many of these post-RGB binaries have B--F type companions.  These systems have similar periods and WD masses to four early type dwarf + He WD systems recently discovered with {\em Kepler} \citep{Rowe:2010,Carter:2011,Breton:2011}.  These objects meet the observable-based criteria we have used to select sdBs, so we include them in the plots below.  However, for clarity we will refer to them as post-RGB stars and any discussion of sdBs below refers only to core-helium burning sdBs.               
   
\autoref{fig:ic_toon} shows the initial primary masses $M_{10}$ and orbital periods $P_{0}$ of the binaries.  The systems that form sdBs through the channels labeled in \autoref{fig:han_toon} come from distinct regions in the $\log P_{0}-M_{10}$ plane.  Again, these regions are schematic and represent the combination of all of our BPS runs with the boundaries changing slightly depending on the parameters used.  The lower limit on $M_{10}$ for sdB progenitors is determined by the 15 Gyr evolution time used in our BPS models.  Had we used a 10 Gyr evolution time, the limit on $M_{10}$ would shift from $\sim 0.8~M_{\sun}$ to $\sim 1~M_{\sun}$.  \autoref{fig:ic_toon} shows this would reduce the number of sdB binaries produced by the RGB CE and the RGB RLOF channels.

Systems that evolve via the HG RLOF channel have primaries with $M_{10} > 2~M_{\sun}$ and orbital periods $P_{0} \la 10$ days.  Binaries with initial periods below the lower bound shown in \autoref{fig:ic_toon} begin RLOF while the primary in on the MS, which results in either the primary losing too much mass to ignite helium in its core or a merger of the two stars.  The post-RGB systems undergo stable RLOF mass transfer that begins while the primary is in the HG, but the envelope is stripped before the core reaches the mass required for helium ignition.  Accordingly, these systems occupy the low-mass, short-period corner of the diagram. For binaries with longer $P_{0}$, the primary does not come into contact with its Roche lobe until it begins to ascend the RGB.  If we allow $q_{\rm crit} = 1.5$ on the RGB, some of these systems can undergo stable mass transfer and form sdBs through the RGB RLOF channel.  At longer periods still, the star will ignite helium and detach from its Roche lobe before the envelope is completely stripped. This forms the long period boundary for RGB RLOF systems.

If we use the \bse~$q_{\rm crit}$ formulation, then systems that begin mass transfer while the primary is on the RGB will enter a CE phase.  We have split this channel into two regions, RGB CE I and RGB CE II.  In the former, $M_{10} \la 1.995~ M_{\sun}$ and these stars only produce sdBs if they can ignite helium below the TRGB.  The primaries in RGB CE II ignite helium non-degenerately on the giant branch.  The lower boundary of the combined RGB CE regions is approximately given by $R_{1}({\rm TRGB}) = R_{\rm L}(P_{0},q_{0} = 1)$, where $R_{1}$ is the radius of the primary and $R_{\rm L}$ is the Roche lobe radius.  Binaries with smaller $q_{0}$ occupy the region above the curve because we have included tidal evolution in all of our models.  Without tidal evolution, the $R_{1}({\rm TRGB}) = R_{\rm L}(P_{0},q_{0} = 1)$ curve would be the upper boundary of this region.  Tidal evolution, which in \bse~drives the binary to a shorter orbital period through spin--orbit angular momentum exchange, is stronger in systems with large $q_{0}$ \citep{Hut:1981}.  Without tidal evolution a binary with $M_{10}\la 1.995$ and $q_{0} < 1$ would need a shorter $P_{0}$ for the primary to fill its Roche lobe near the TRGB.  We note that the upper boundary of the RGB CE I region in our models is flat because we require $M_{2} > 0.1~M_{\sun}$.
\begin{figure*}
	\centering
	\includegraphics[width=0.9\textwidth]{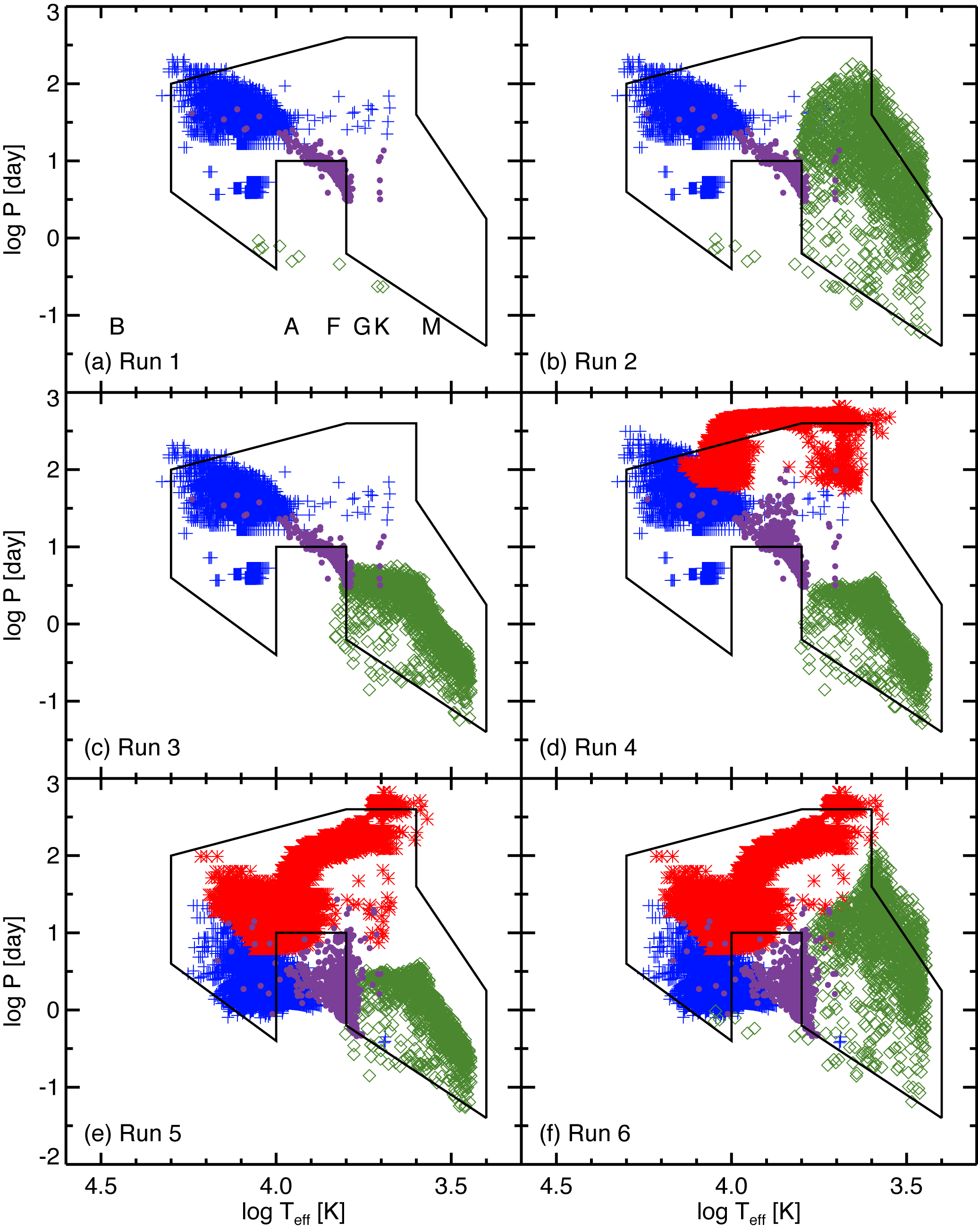}
	\caption{$\log P$ vs $\log T_{\rm eff}$ of the MS companion for 6 \bse~runs. In each panel, $\gamma = 1$ and \ace~= 0.75.  Blue crosses and red asterisks represent systems that underwent stable mass transfer beginning while the primary was in the HG or RGB, respectively. Systems that underwent CE evolution that began while the primary is on the RGB are plotted with green diamonds.  Binaries that contain a post-RGB object instead of a core helium burning sdB are plotted as purple circles.  The black lines give a rough outline of the region occupied by systems from H03.  Panel (a) shows our fiducial run using all \bse~default values, Run 1.  Panel (b) shows Run 2, identical to Run 1 but setting $f_{\rm He}$ = 0.95.  Panel (c) shows Run 3, identical to run 2 but using the  LVK11 envelope binding energies.  Panel (d) shows Run 4, identical to Run 3, but setting $q_{\rm crit} = 1.5$ on the RGB.  Panel (e) shows Run 5, identical to Run 4 but setting $f_{\rm RLOF}=0.5$.  Panel (f) shows Run 6, the same as Run 5 but with the \bse~envelope binding energies.  The temperature regimes for spectral types B--M are shown in panel (a). \label{fig:han1}}
\end{figure*}
\begin{figure*}
	\centering
	\includegraphics[width=0.9\textwidth]{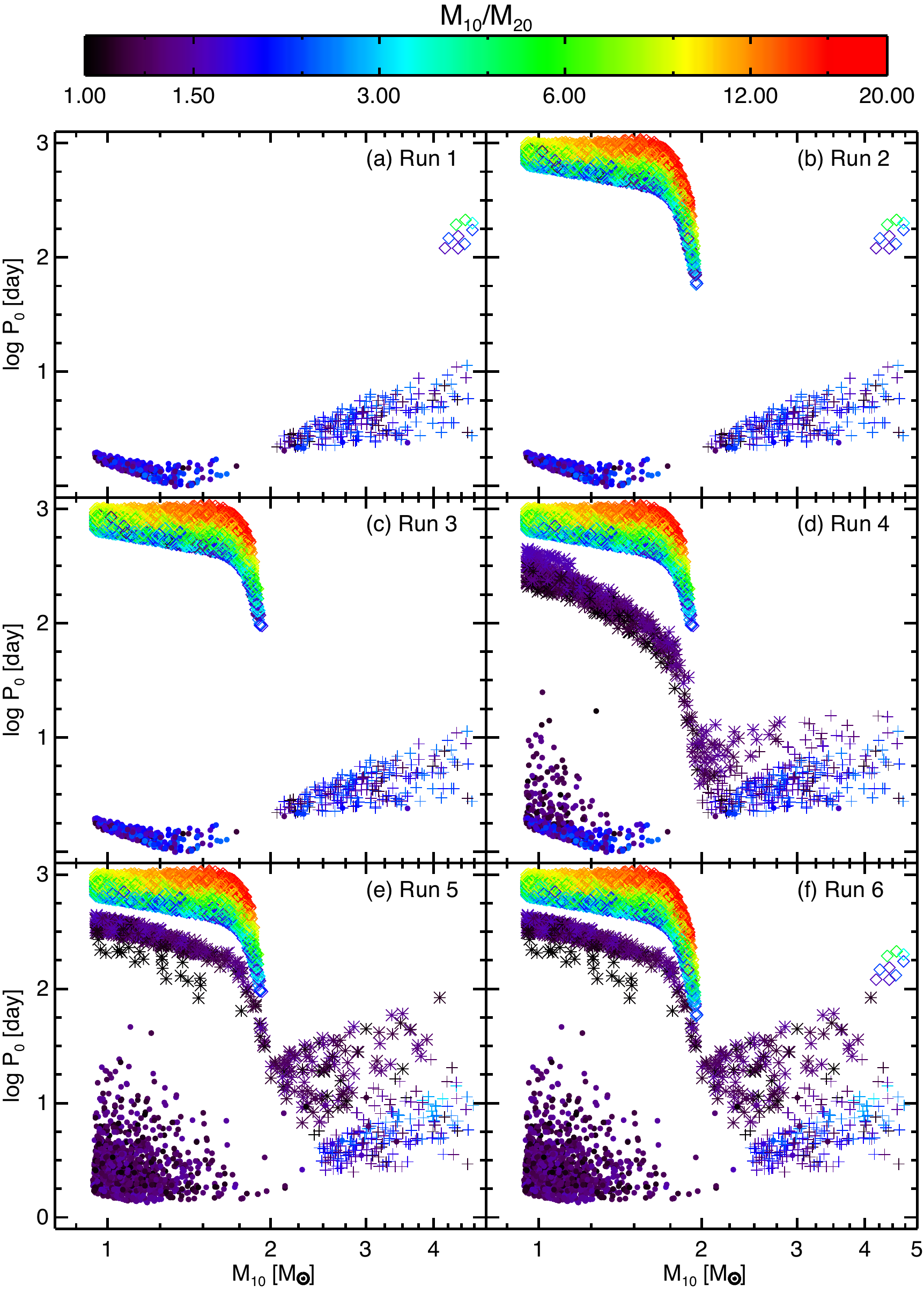}
	\caption{$\log P_{0}$ vs $M_{10}$ (initial orbital period and primary mass, respectively) for the \bse~runs shown in \autoref{fig:han1}.   Crosses and asterisks represent systems that will undergo stable mass transfer beginning while the primary is in the HG or RGB, respectively. Systems that will undergo CE evolution that begins while the primary is on the RGB are plotted with diamonds.  The circles are systems that produce a post-RGB object instead of a core helium burning sdB.  The color of each point shows the system's initial mass ratio $q_{0}$ (see color bar at top of figure). (a) Run 1, (b) Run 2, (c) Run 3, (d) Run 4, (e) Run 5, and (f) Run 6.\label{fig:ic1}}
\end{figure*}
\begin{figure*}
	\centering
	\includegraphics[width=0.9\textwidth]{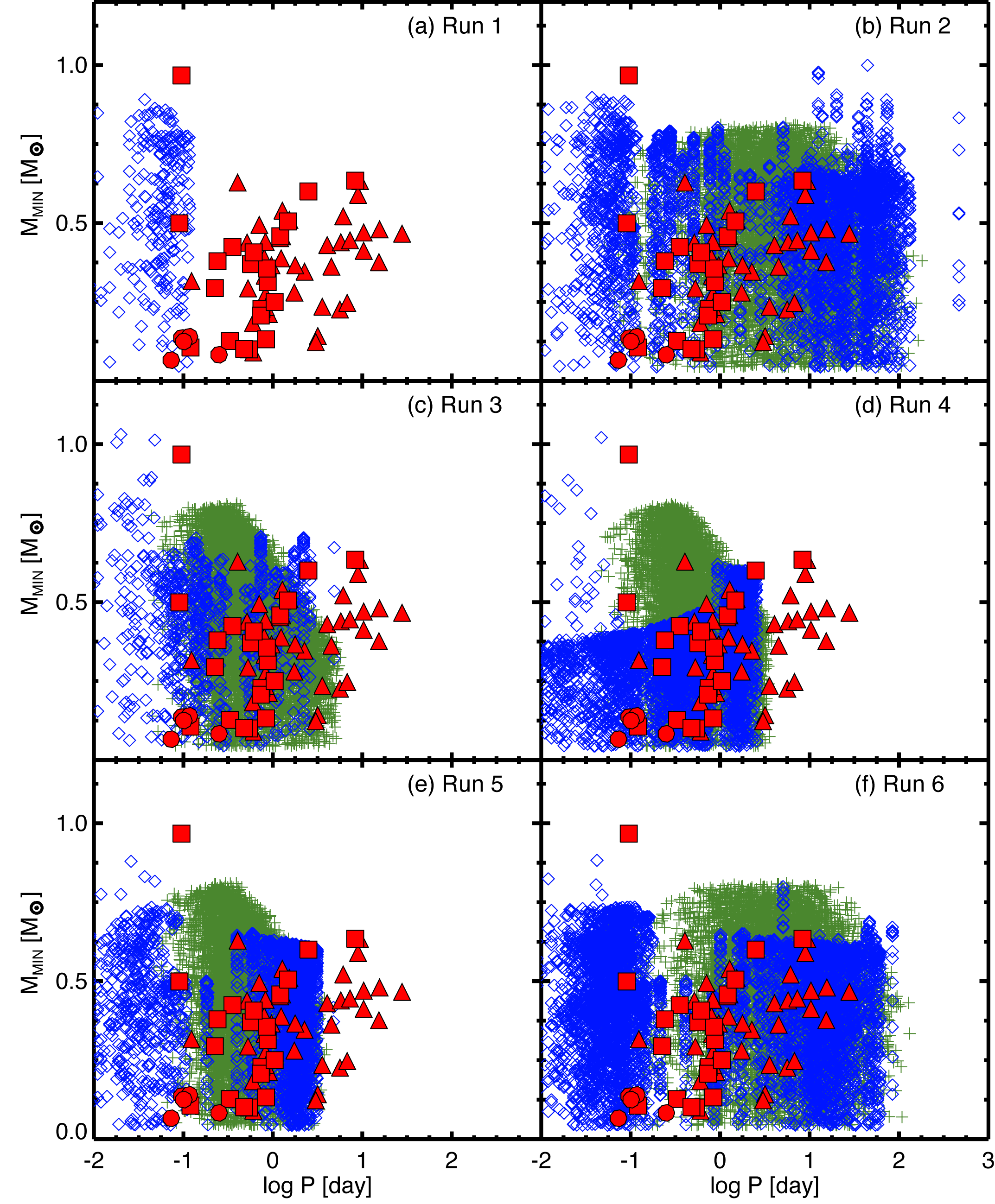}
	\caption{Simulated minimum companion mass (see \autoref{sec:parms}) vs $\log P$ for the \bse~runs shown in \autoref{fig:han1}.  Only sdB + WD (blue diamonds) and sdB + M dwarf (green crosses) binaries are plotted.  The ``towers'' of points at constant $P$ are due to a single, resampled binary ``observed'' at different orientation angles. The red symbols are an observed sample of sdB binaries compiled from \citet{Morales-Rueda:2003} and \citet{Copperwheat:2011}.  The squares are sdBs with WD companions, the circles are sdBs with M dwarf companions, and the triangles are systems with unknown companion type.  (a) Run 1, (b) Run 2, (c) Run 3, (d) Run 4, (e) Run 5, and (f) Run 6.\label{fig:wd1}}
\end{figure*}
\begin{figure*}
	\centering
	\includegraphics[width=0.9\textwidth]{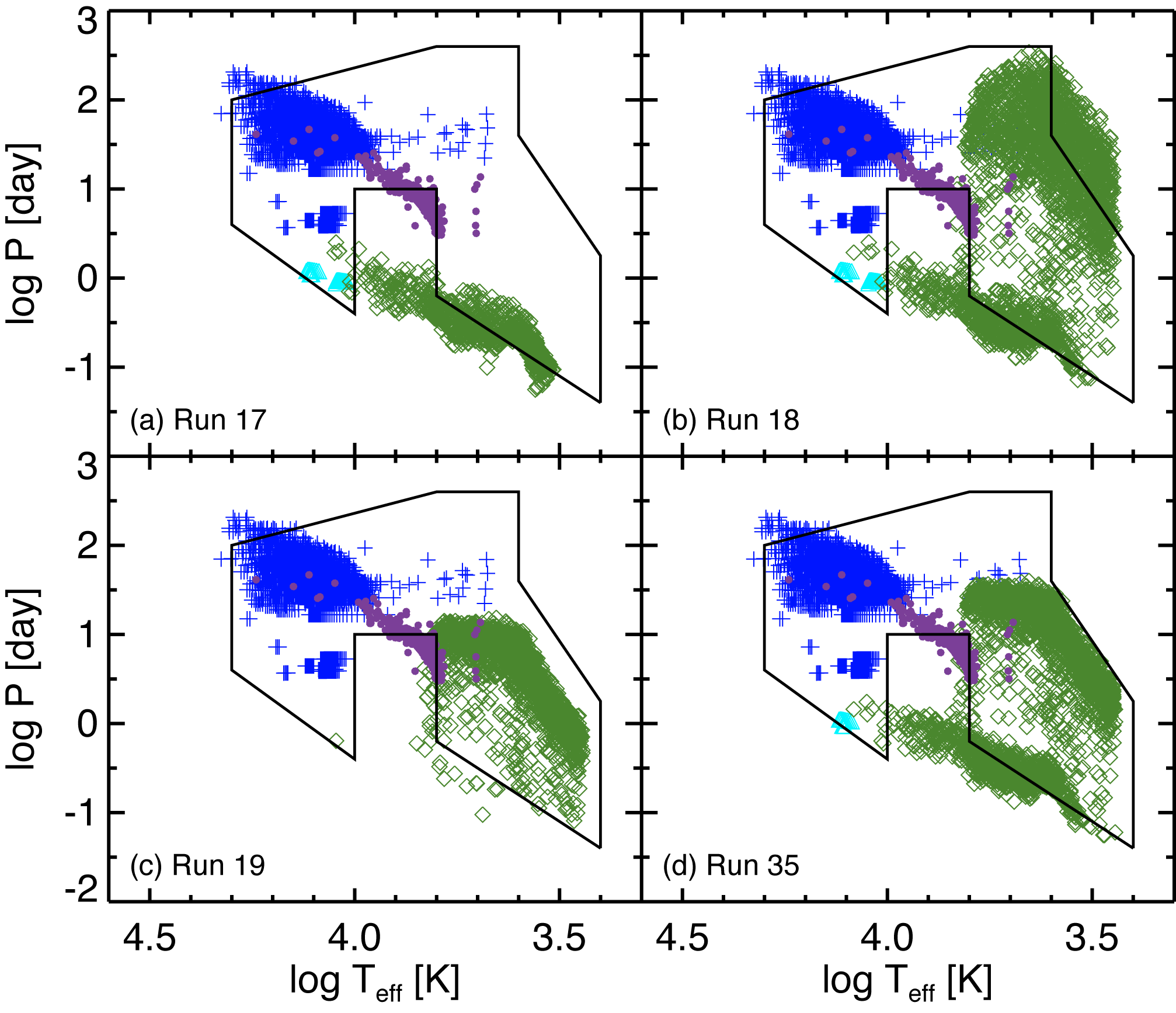}
	\caption{Same as \autoref{fig:han1} with, additionally, light blue triangles representing systems that underwent CE evolution beginning while the primary was in the HG for (a) Run 17 showing the effect of setting $a_{\rm CE} = 1.5$, (b) Run 18 showing the effect of setting $f_{\rm He} = 0.95$, (c) Run 19 showing the effect of using the  LVK11 envelope binding energies, and (d) Run 35 showing the effect of increasing $a_{\rm CE}$ to 3.0.\label{fig:han2}}
\end{figure*} 
\begin{figure*}[!]
	\centering
	\includegraphics[width=0.9\textwidth]{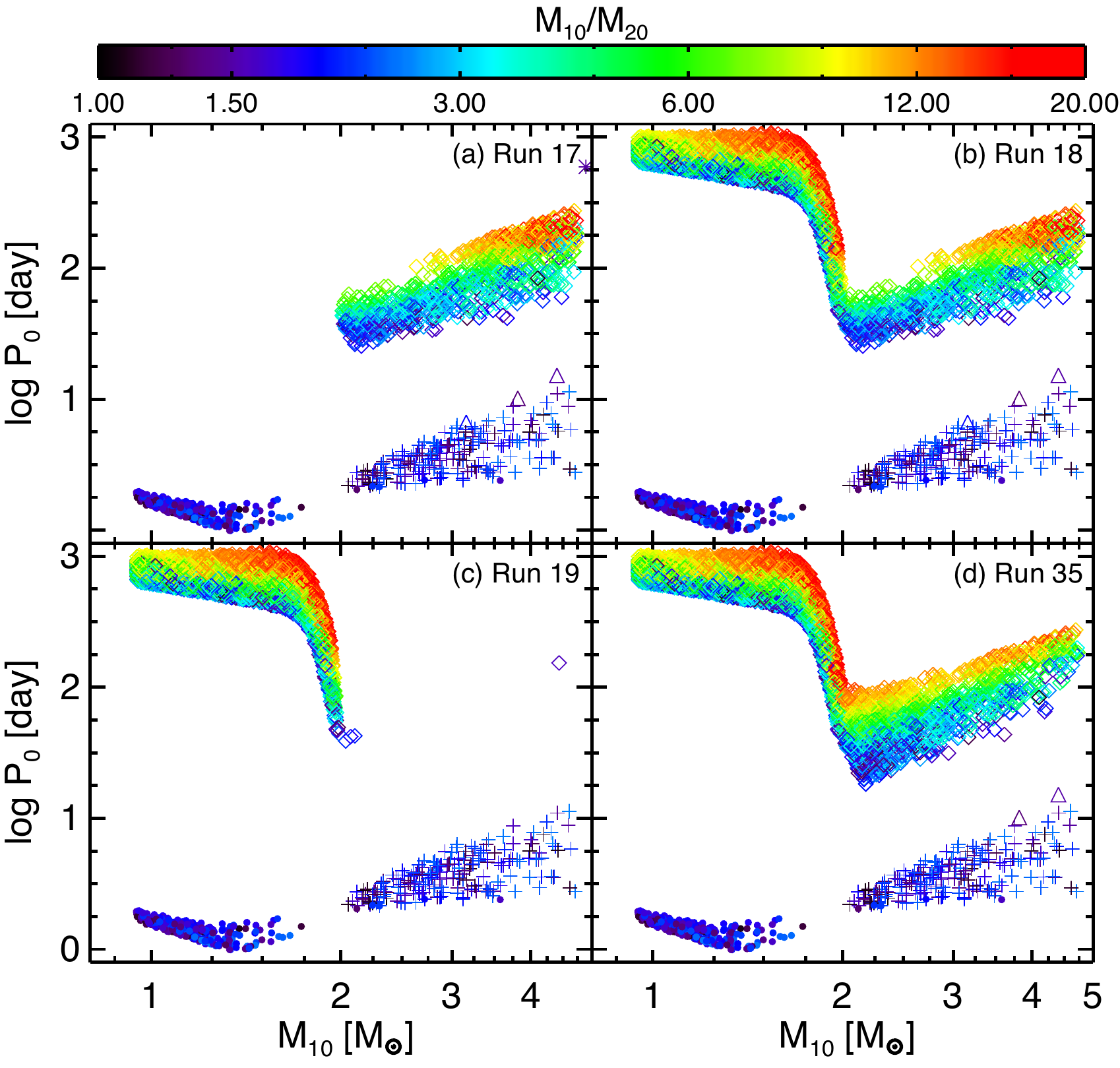}
	\caption{Same as \autoref{fig:ic1} with each panel corresponding to the \bse~runs shown in \autoref{fig:han2}.   (a) Run 17, (b) Run 18, (c) Run 19, and (d) Run 35.\label{fig:ic2}}
\end{figure*}
\begin{figure*}
	\centering
	\includegraphics[width=0.9\textwidth]{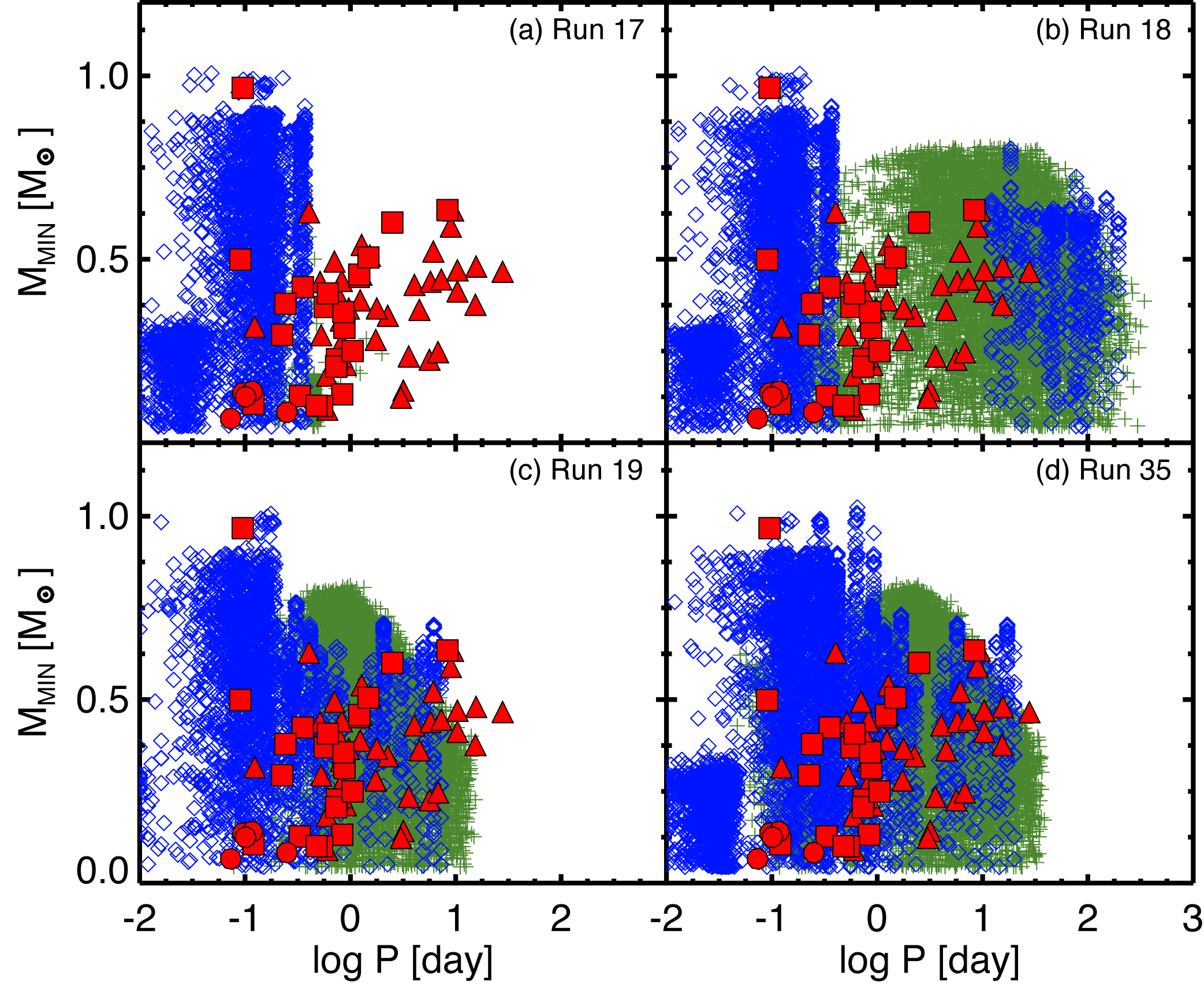}
	\caption{Same as \autoref{fig:wd1} with each panel corresponding to the \bse~runs shown in \autoref{fig:han2}.  (a) Run 17, (b) Run 18, (c) Run 19, and (d) Run 35.\label{fig:wd2}}
\end{figure*}

To evaluate each set of parameters we compared the modeled populations with the observed properties of sdB binaries presented in \citet{Morales-Rueda:2003} and \citet{Copperwheat:2011}.  To replicate selection biases in the observed sample, we  only compare the sdB + WD and sdB + M dwarf binaries predicted by our BPS models.  The observed sample is highly incomplete so we can only test whether the BPS populations span the range of observed orbital periods and companion masses.  Since we cannot compare the density of systems in the observed and synthesized sample, for each model we compute a model suitability parameter $\rho$, which is the fraction of observed systems that lie within boxes that enclose the modeled systems in companion mass--orbital period space.  In some runs, the sdB + WD binaries exhibit a bimodal period distribution with a large gap between the short and long period systems.  In cases where this gap in $\log P \ge 0.5$ dex we draw separate boxes around the long and short period distributions.  Furthermore, because the companion type is known for 26 binaries in the observed sample, we draw separate boxes around the sdB + WD and sdB + M dwarf binaries produced in our BPS runs and require that these observed systems fall within the appropriate box.  The remaining 40 systems, for which the companion type is not known, are included in $\rho$ if they lie in either box.  We list $\rho$ for each run in \autoref{table}.       

\subsection{Parameter Variations}
\label{sec:parms}

To illustrate how parameter variations impact the population of sdB binaries, we will move from one set of parameters that uses the \bse~defaults to another set that is comparable to that of run 2 in H03 with successive, cumulative changes.  In all of these runs, we used the same common envelope ejection efficacy,  $\alpha_{\rm CE} = 0.75$.  Furthermore, the specific angular momentum carried away by mass lost from the system during stable mass transfer was also the same for each of these runs; the material lost during stable RLOF carried away the specific orbital angular momentum of the system ($\gamma = 1$). 

\autoref{fig:han1}a shows the population predicted by Run 1, which we will use as our fiducial run.  The solid black lines bound the region populated by the H03 best fit models, and are shown for comparison.  A single initial binary can produce multiple points on these diagrams in proportion to the time it spends in the sdB ``box'' because of the resampling scheme used to generate a present day population.  Sequences of symbols for the same initial system lie along that binary's evolutionary track and may be roughly horizontal if the companion evolves across the HG or vertical due to period evolution through magnetic braking or tidal exchange of spin and orbital angular momentum.  We warn that the density of systems in a particular region cannot be read from these plots because several binaries are plotted on top of one another.  Marginal distributions are discussed in \autoref{sec:dists} and the number of systems formed thorough each channel is given in \autoref{table}.  

In Run 1, almost all of the sdBs with hydrogen burning companions were formed through stable RLOF initiated while the primary was in the HG.  The primaries in these systems had initial masses $M_{10} > 2~M_{\sun}$, which can be seen in \autoref{fig:ic1}a which shows the initial mass of the primary $M_{10}$, initial period $P_{0}$, and mass ratio $q_{0}$ for the sdB + MS binaries.  Their evolution was similar to that of Algol, the stable mass transfer eventually leading to mass ratio reversal and an increase in the orbital separation.  This resulted in sdBs with early type companions in systems with $P > 10$ days.  In some cases, the companions have evolved into or across the Hertzsprung gap.  In this run, there are nine sdBs that formed as the result of a CE phase that began while the primary was on the giant branch.  These systems all have $P < 1$ day.        
\begin{figure*}
	\centering
	\includegraphics[width=0.9\textwidth]{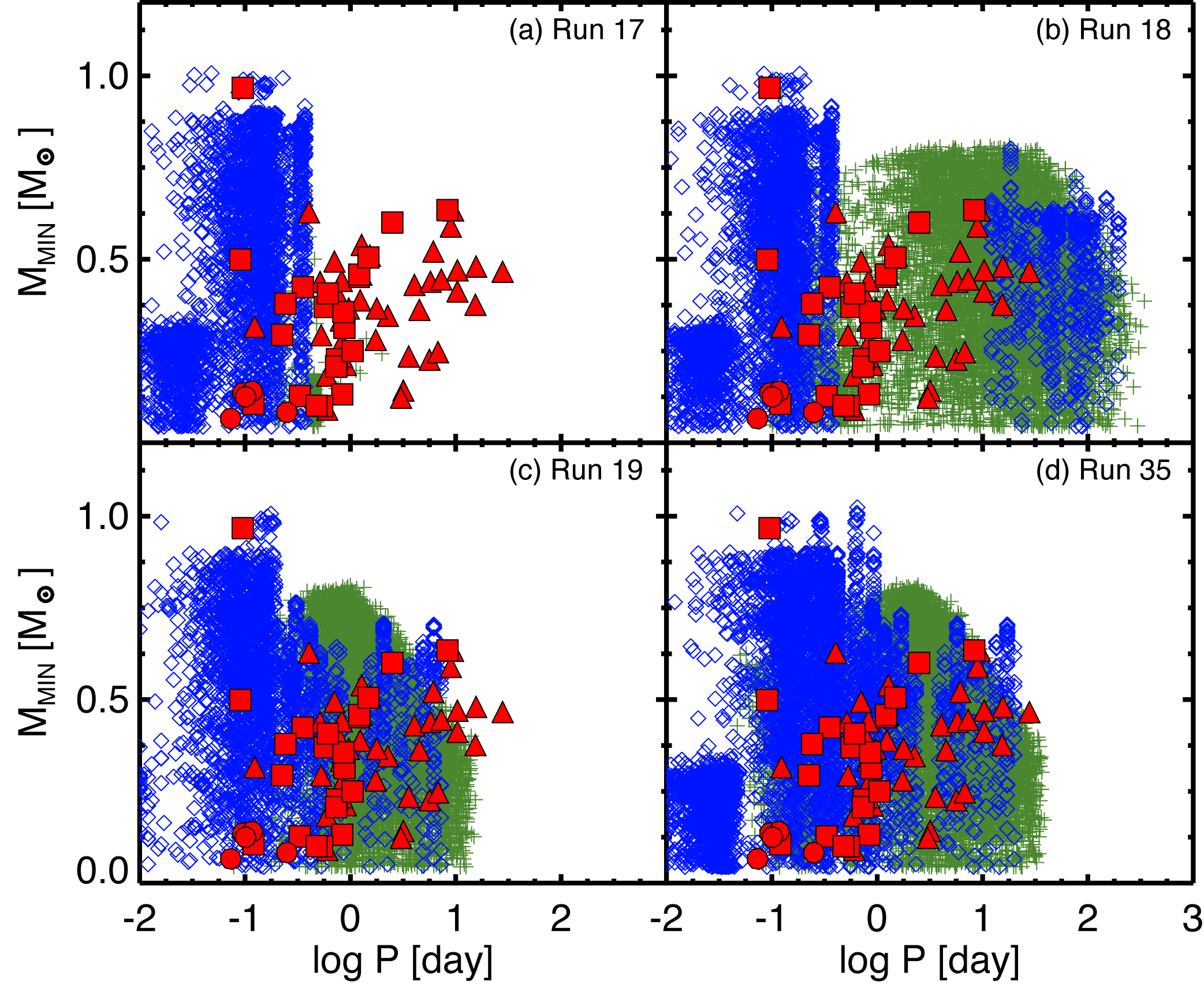}
	\caption{Same as \autoref{fig:wd1} with each panel corresponding to the \bse~runs shown in \autoref{fig:han2}.  (a) Run 17, (b) Run 18, (c) Run 19, and (d) Run 35.\label{fig:wd2}}
\end{figure*}
\begin{figure*}
	\centering
	\includegraphics[width=0.9\textwidth]{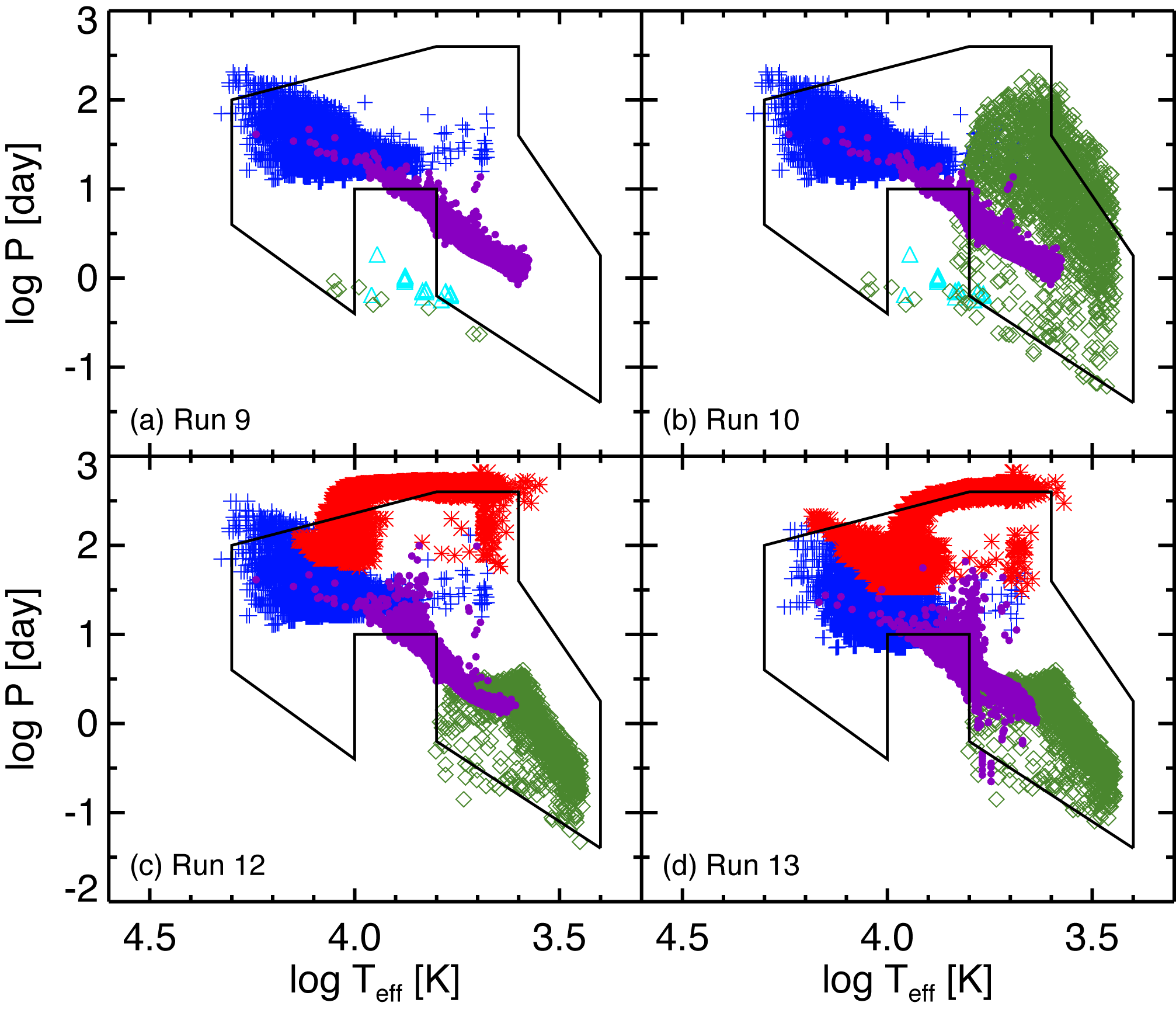}
	\caption{Same as \autoref{fig:han1} for (a) Run 9 showing the effect of allowing the ejected material to carry away the specific angular momentum of the donor ($\gamma = -1$), (b) Run 10 showing the effect of setting $f_{\rm He} = 0.95$, (c) Run 12 showing the effects of setting $q_{\rm crit} = 1.5$ and using the  LVK11 envelope binding energies, and (d) Run 13 showing the effect of setting $f_{\rm RLOF} = 0.5$.\label{fig:han3}}
\end{figure*}
\begin{figure*}
	\centering
	\includegraphics[width=0.9\textwidth]{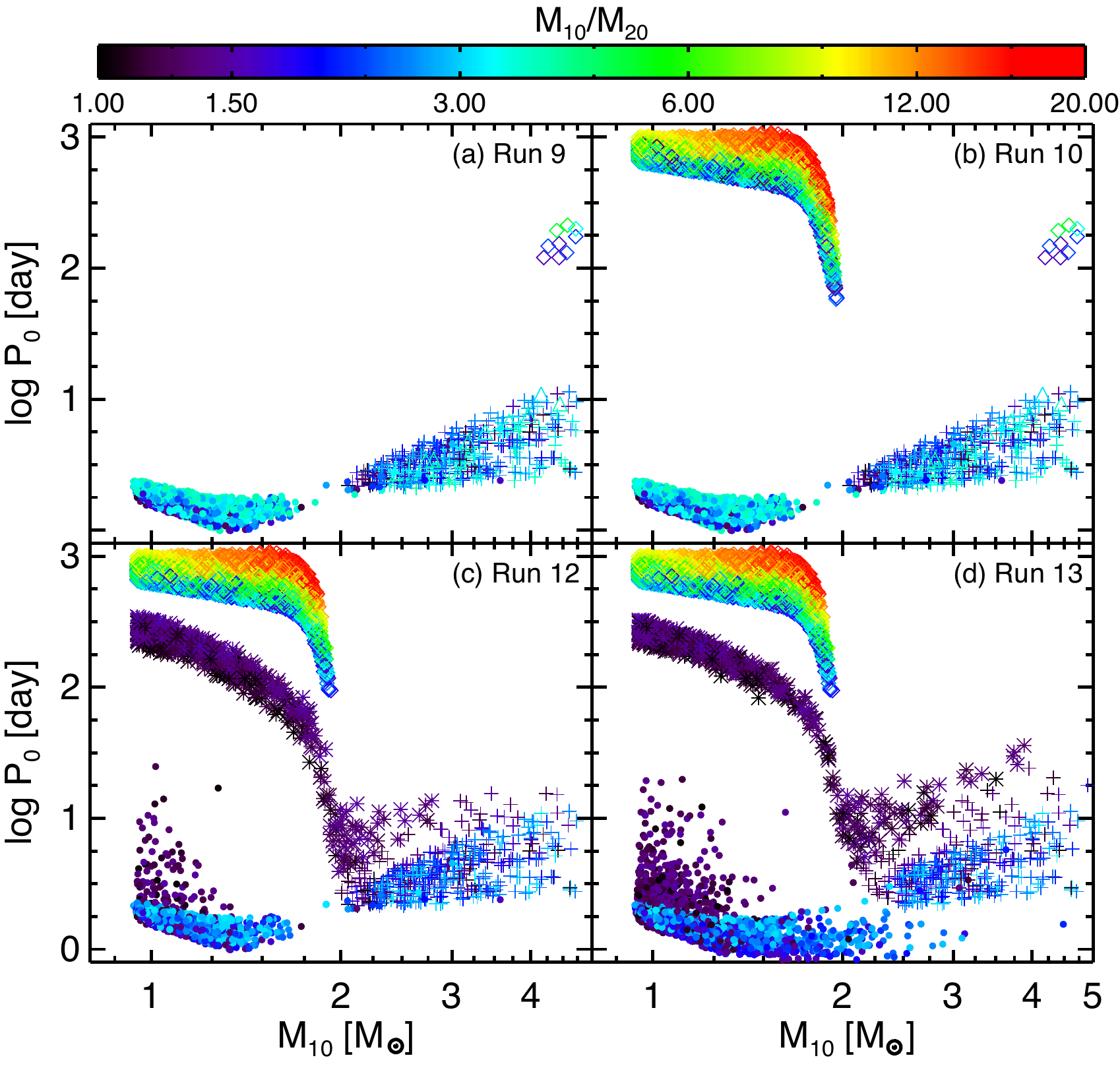}
	\caption{Same as \autoref{fig:ic1} with each panel corresponding to the \bse~runs shown in \autoref{fig:han3}.   (a) Run 9, (b) Run 10, (c) Run 12, and (d) Run 13.\label{fig:ic3}}
\end{figure*}
\begin{figure*}
	\centering
	\includegraphics[width=0.9\textwidth]{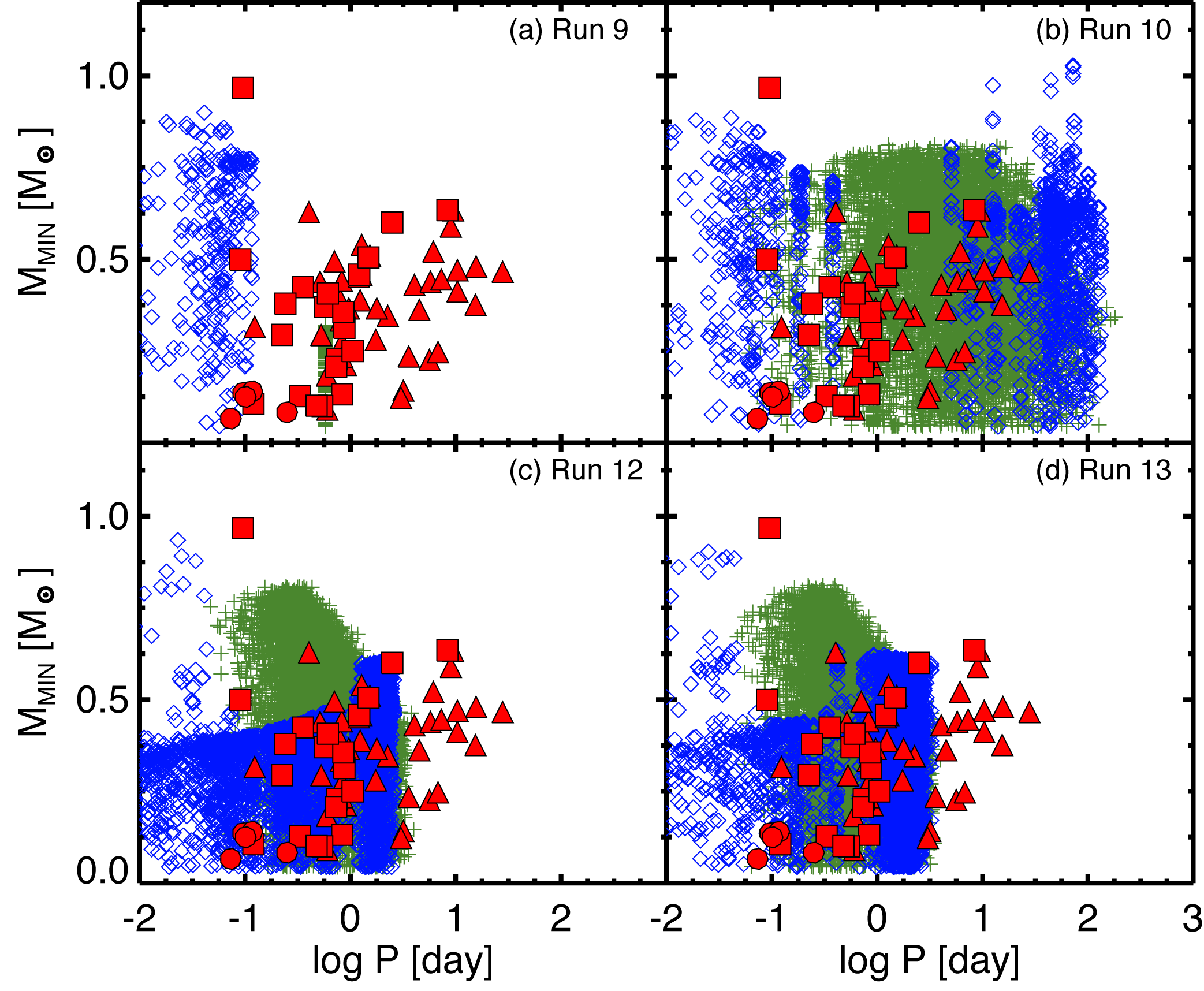}
	\caption{Same as \autoref{fig:wd1} with each panel corresponding to the \bse~runs shown in \autoref{fig:han3}.  (a) Run 9, (b) Run 10, (c) Run 12, and (d) Run 13.\label{fig:wd3}}
\end{figure*}

When we allowed stars to ignite helium if their cores had reached a mass $m_{\rm c} = 0.95 m_{\rm c}({\rm TRGB})$ before the envelope was lost, far more sdBs were formed during CE evolution that began while the primary was on the RGB (Run 2, \autoref{fig:han1}b).  The vast majority of the sdBs formed through CE evolution have G, K, or M type companions and the orbital periods of these binaries range from 0.05--220 days.   \autoref{fig:ic1}b shows that primaries in these systems (green) had initial masses in the range $0.8 - 2~M_{\sun}$ and that the initial periods of these systems were in the range 60--1100 days.  In Run 1, these systems survived the CE phase, but the primary never ignited helium and became a helium WD without passing through the sdB ``box." 

In Run 3,  we have used the envelope binding energies computed by  LVK11 (\autoref{fig:han1}c).  These more tightly bound envelopes resulted in a much narrower range of orbital periods for post CE systems, none of which exceed 10 days.  Additionally, some systems were unable to eject the more tightly bound envelope and merged without ever producing an sdB.

Setting $q_{\rm crit} = 1.5$ on the RGB and  3.2 in the HG (Run 4, \autoref{fig:han1}d) allowed many systems to avoid the CE phase and undergo stable RLOF mass transfer on the RGB.  These systems produce binaries consisting of a sdB with an A to G type MS companion and orbital periods in the range 55--660 days. As was the case with the systems formed by stable RLOF mass transfer initiated in the HG, some of the companions have evolved into later type giants.  There was also a slight decrease in the number of short period sdB + F, G, or K type dwarf systems because of the higher value of $q_{\rm crit}$.  With $q_{\rm crit} = 1.5$, these systems underwent stable RLOF mass transfer that did not completely remove the envelope and never produced an sdB.  Instead, they reach He ignition mass and detach from the Roche lobe before the entire envelope is transferred; they become horizontal branch stars.  

We considered a fixed fraction of mass loss during stable RLOF mass transfer, $f_{\rm RLOF}=0.5$ (Run 5,  \autoref{fig:han1}e).  This resulted in far more mass loss during stable RLOF mass transfer than in the previously discussed runs.  This non-conservative mass transfer had two effects on the resulting population.  First, the mass lost from the system carried away angular momentum which drove the binaries to shorter periods. Second, since less material was accreted by the secondary, the companions have lower masses and, thus, lower $T_{\rm eff}$ than they did in the nearly conservative case. 

Finally, in \autoref{fig:han1}f, we show the population predicted by Run 6, in which we used the default \bse~values for the envelope binding energy.  It is useful to compare the post-CE populations of Run 6 and Run 2 (\autoref{fig:han1}b).  Increasing $q_{\rm crit}$ has depleted the high $\log P$, high $\log T_{\rm eff}$ corner of the post-CE population.       

Runs 1--6 produced a diverse set of populations of sdB + MS binaries. The orbital periods were in the range 0.04--663 days and companions were of every spectral type from B to M.  Furthermore, it was possible to create these vastly different populations while holding $\alpha_{\rm CE}$ and $\gamma$ constant.  In \autoref{fig:wd1} we compare the sdB + WD and sdB + M dwarf binaries predicted by our BPS models with the observed sdB binaries from  \citet{Morales-Rueda:2003} and \citet{Copperwheat:2011}.  To facilitate the comparison with observed sdB binaries, we have computed a minimum mass for the WD or M dwarf companions:  $M_{2} \sin^3 i$ is ``observed'' for each theoretical present day system, via the mass function $f(M)=P K_{1}^3/2 \pi G = M_2^3 \sin^3i/(M_1+M_2)^2$; where $K = 2\pi a_1\sin i/P$ with $i$ chosen so the distribution of $\cos i$ is uniform and the other quantities
known from model output.  Assuming $M_1=0.48~M_{\sun}$ then allows us to infer  $M_2 \sin^3 i$, and setting $\sin i = 1$ gives $M_{\rm min}$.  Again, a single initial binary is responsible for several points on this diagram.  In \autoref{fig:wd1} the towers of points at nearly constant $P$ correspond to a single, resampled binary evolution track with an independent inclination for each point.  In many of our BPS runs, one or two sdB + neutron star binaries were formed.  We included these binaries with the sdB + WD systems because the two are indistinguishable in terms of the simulated $M_{\rm min}$.     

The fraction of observed systems reproduced by each run is listed in column (19) of \autoref{table}.  Clearly, Runs 1, 3, 4, and 5 are unable to reproduce many of the observed systems.  The populations produced by Runs 2 and 6, on the other hand, are consistent with the observed population.  However, despite the similarity amongst the sdB + WD and sdB + M dwarf  populations in Run 2, Run 6, and the observed sample, the two models make strikingly different predictions for the population of sdB + MS binaries, as can be seen in \autoref{fig:han1}b and \autoref{fig:han1}f.

In each run, $3-11\%$ of the sdB + WD systems had previously been sdB + MS binaries.  In these systems the primary becomes an sdB through an episode of stable mass transfer.  Eventually this sdB evolves into a WD and after the secondary begins to ascend the RGB the system undergoes CE evolution resulting in a sdB + WD system.  The period and mass distributions of the sdB + WD binaries that form through this channel overlap with those of the sdB + WD binaries formed in systems that have not previously undergone an sdB phase, which makes it difficult to distinguish between the two subpopulations.     

\subsubsection{Influence of \ace}
We demonstrate the effect of varying $\alpha_{\rm CE}$ in Figures 5-8, again with successive, cumulative changes to the parameters.  \autoref{fig:han2}a shows Run 17, which was the same as our fiducial run, shown in \autoref{fig:han1}a, except we increased \ace~to 1.5.  In this case, the orbital energy was efficiently transferred to the CE as the stars spiraled towards one another and many more systems were able to eject the CE before merging.  The result is a band of systems with $P \la 5$ days with companions spanning all types from early A to M.  \autoref{fig:ic2}a shows that these post-CE systems occupy a different region of parameter space than most of the post-CE systems seen in Runs 1-6, namely that $M_{10} \ga 2~M_{\sun}$ for these systems.  The primaries in these binaries were able to ignite helium non-degenerately in their cores, despite the loss of the H envelope in the CE phase.   

For Run 18, we kept \ace~= 1.5 and set $f_{\rm He} = 0.95$.  This resulted in a large number of additional post-CE systems, most of which had longer periods than the post-CE systems produced in Run 17.  These systems have longer periods because their formation required that the primary did not fill its Roche lobe until it was near the TRGB, otherwise the CE phase would have begun prior to the core reaching the mass required for helium ignition.  This, in turn, required binaries with large initial separations, in order to increase the size of the Roche lobe.  Comparing panel (a) of \autoref{fig:ic1} with panels (a) and (b) of \autoref{fig:ic2} one can clearly see the different regions of initial parameter space that produced post-CE systems as a result of increasing \ace~(the band of systems with $M_{10}>2~M_{\sun}$) and as a result of reducing $f_{\rm He}$ (the systems with $M_{10} < 2~M_{\sun}$). 

We used the  LVK11 binding energies in Run 19, shown in \autoref{fig:han2}c.  Even though the orbital energy was efficiently transferred to the envelope in this run, the higher envelope binding energies in Run 19 resulted in most of the systems being driven to merger, which in Run 18 ignited helium non-degenerately and survived the CE phase.  In \autoref{fig:han2}d we show Run 35, which is identical to Run 19 except we have set \ace~= 3.0.  Allowing the orbital energy to be transferred to the envelope even more efficiently allows systems to avoid mergers, even with the increased binding energies of  LVK11.  

\autoref{fig:wd2} shows that Run 35 is the only one of these four models that is consistent with observations.  The short period binaries in Run 18 do span the observed range of period and mass, however the large gap in the sdB + WD period distribution is problematic.  Several of the observed sdB binaries known to have WD companions  fall within the gap, which makes the population inconsistent with the sample.      

\begin{figure}
	\centering
	\includegraphics[width=0.45\textwidth]{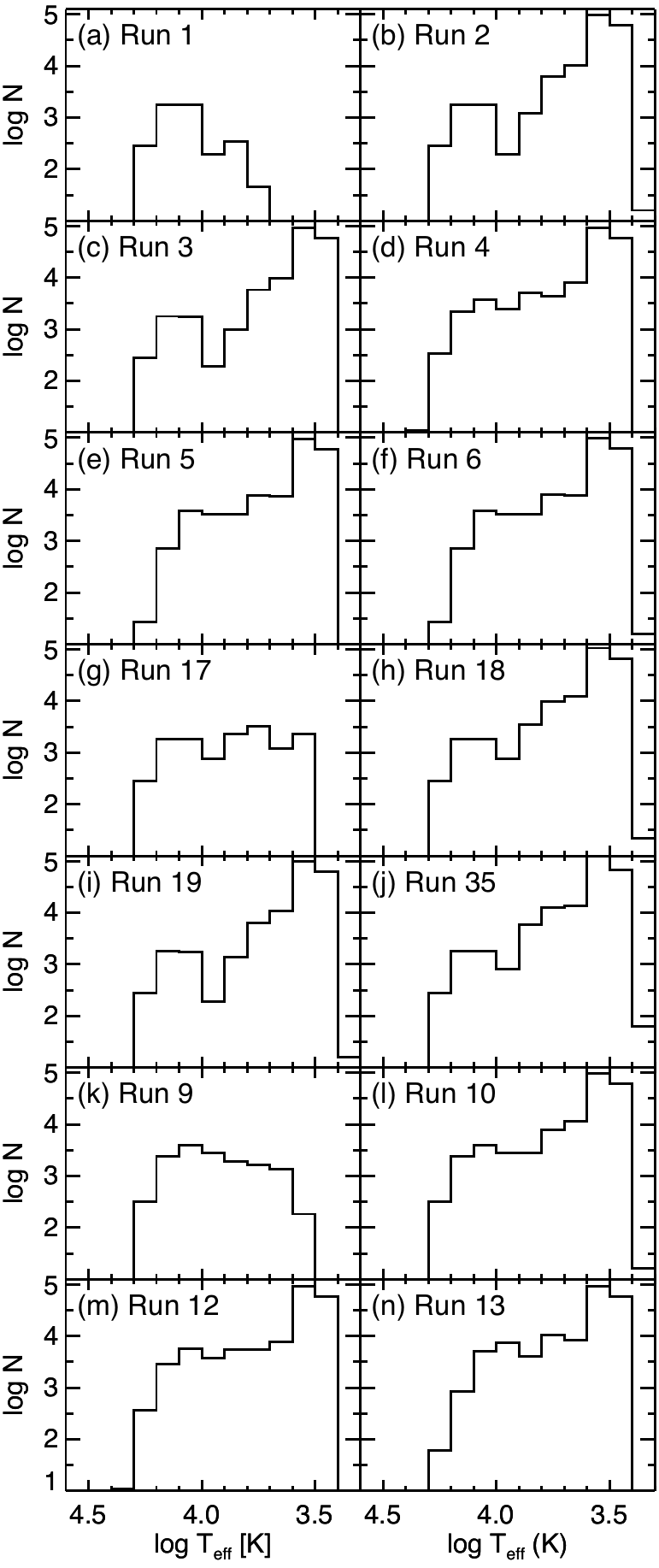}
	\caption{Histograms of the companion $\log T_{\rm eff}$ distribution for the runs shown in Figures 1-11. Both true sdB and post-RGB binaries are included.\label{fig:teff_hist}}
\end{figure}
\begin{figure}
	\centering
	\includegraphics[width=0.45\textwidth]{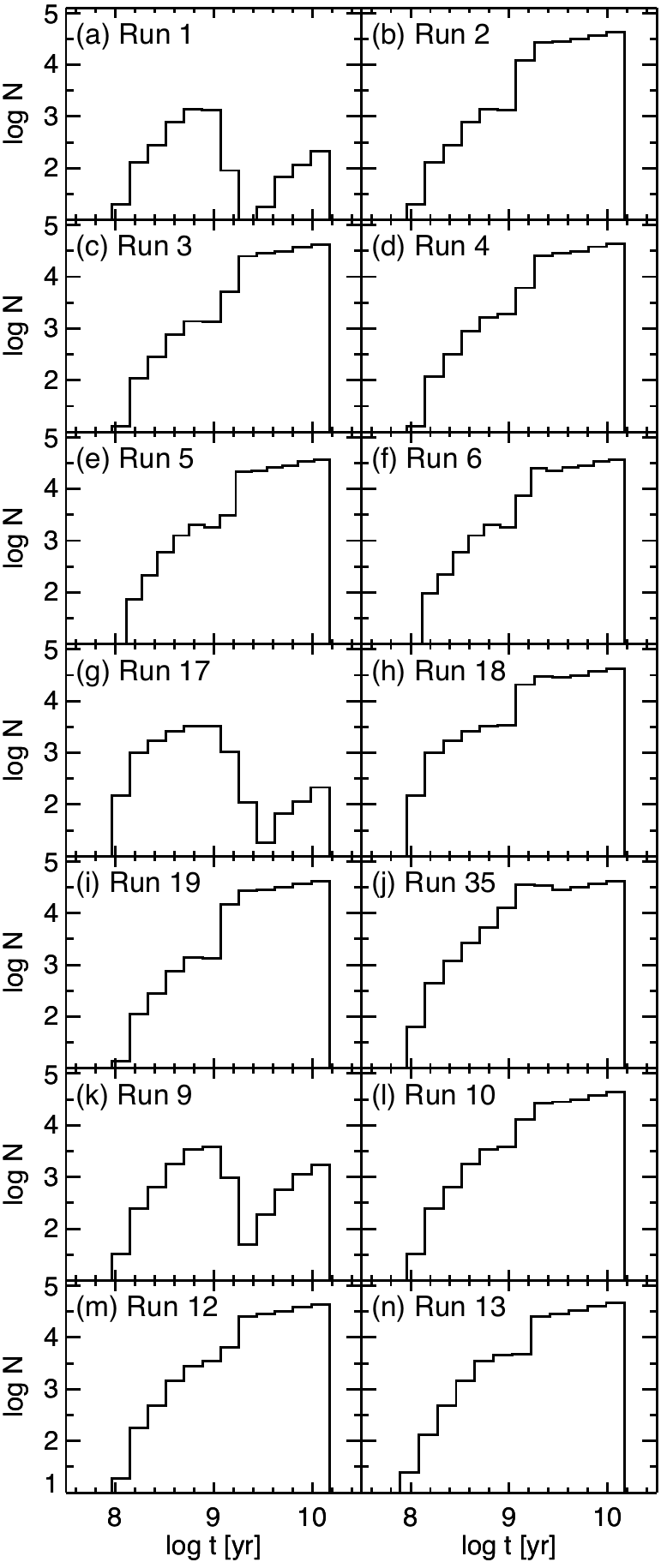}
	\caption{Histograms of the age of the sdB + MS and post-RGB + MS binaries for the runs shown in Figures 1-11. \label{fig:age_hist}}
\end{figure}
\begin{figure}
	\centering
	\includegraphics[width=0.45\textwidth]{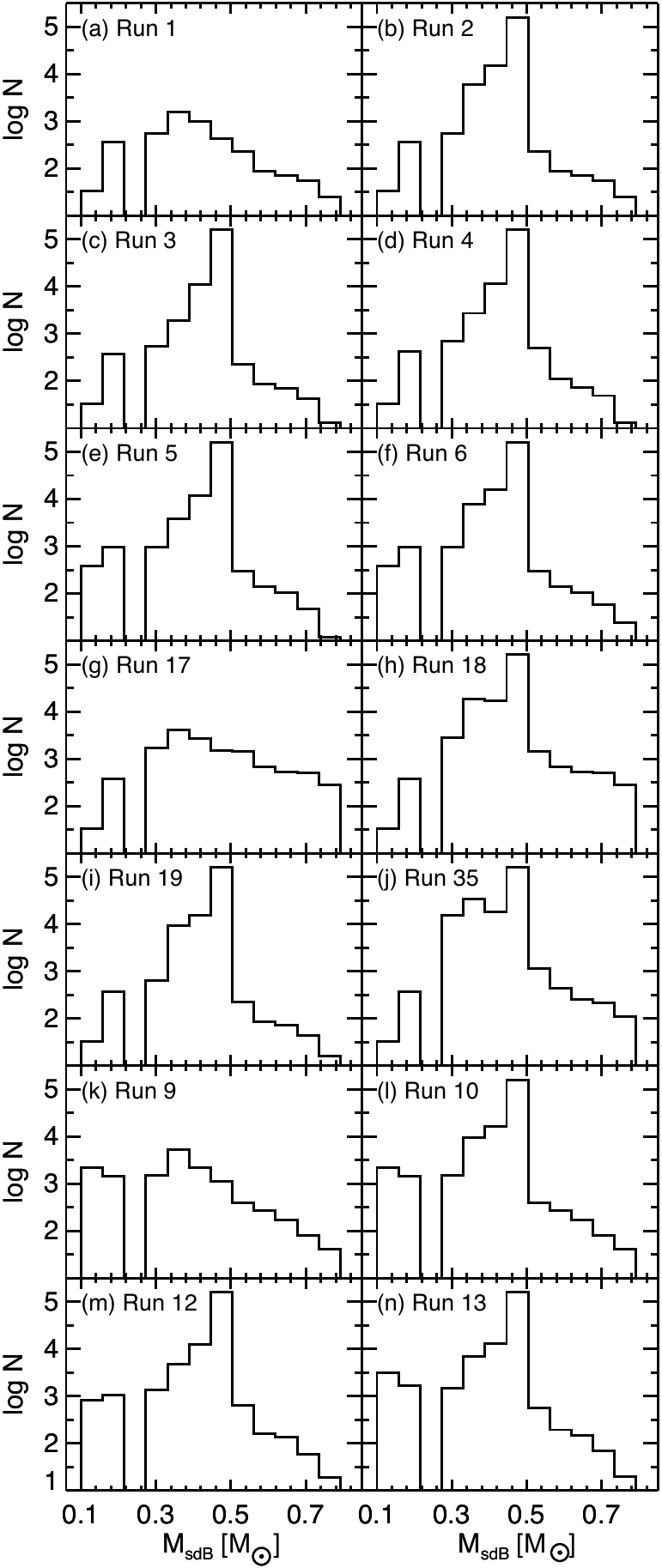}
	\caption{Histograms of $M_{\rm sdB}$ for the runs shown in Figures 1-11.  Both true sdB and post-RGB binaries are included.\label{fig:msdb_hist}}
\end{figure}

\begin{figure}
	\centering
	\includegraphics[width=0.47\textwidth]{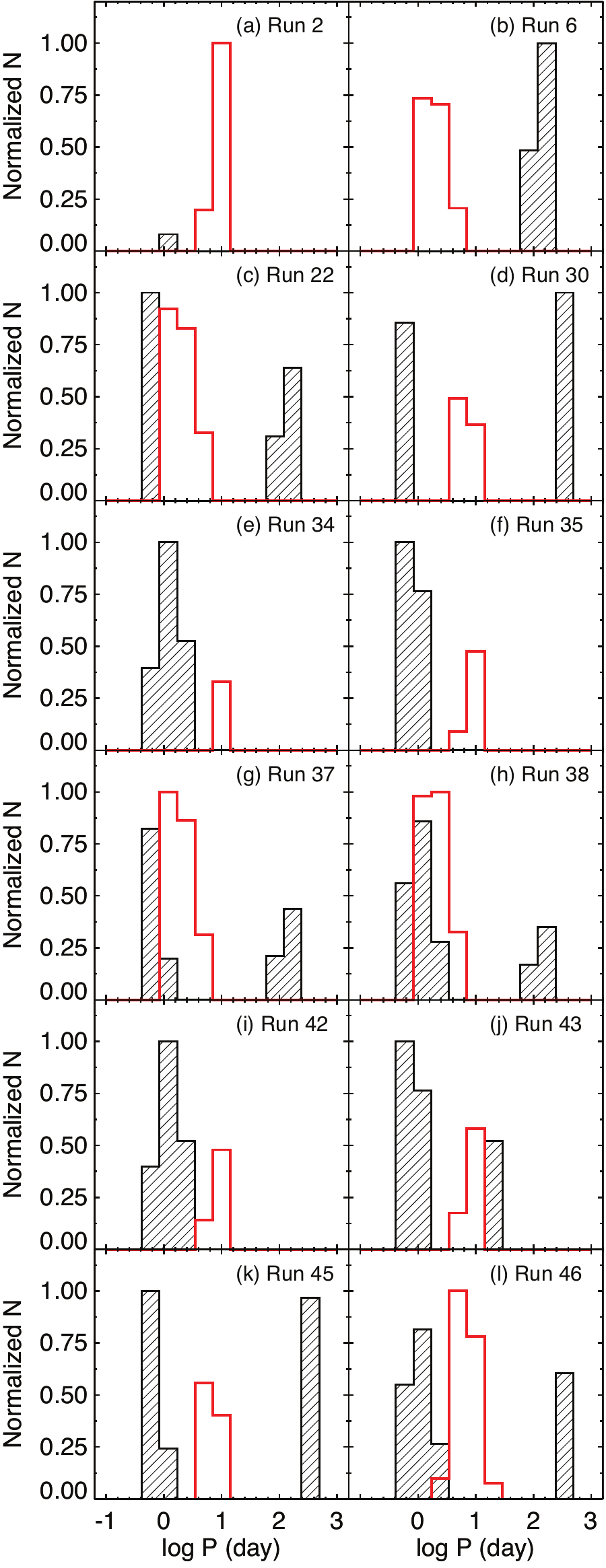}
	\caption{Histograms of the orbital periods of helium burning sdB + early F binaries ($7350 > T_{\rm eff} > 6700$, hatched) from the ``focused'' BPS runs and post-RGB + early F binaries (red) from the twelve runs with a model suitability parameter $\rho > 0.9$ (see \autoref{section:gen}).  The post-RGB histograms have been rescaled to be in proper proportion to the ``focused'' models and we have fixed the larger of the two density distributions maxima to unity, to emphasize the relative differences between distributions on the same scale.\label{fig:fhist}}
\end{figure}

\subsubsection{Influence of $\gamma$}
Figures 9-12 illustrate how changing the specific angular momentum carried away by material lost from the system during mass transfer affects the population of sdB + MS binaries.  In each of the runs shown in these figures, the material lost from the system carries away the specific angular momentum of the donor $a_{\rm d}^{2}\Omega_{\rm Orb}$.  The angular momentum lost per unit mass in these runs is at least a factor of four lower (the primary is the donor so $a/a_{1} = (q + 1)$ and $q \ge 1$) than in the previously discussed runs.  \autoref{fig:han3}a shows Run 9, which used the same parameters as our fiducial run except for angular momentum loss.  Reducing the amount of angular momentum carried away by ejected material resulted in the formation of sdB + A or F dwarf binaries with $P > 10$ days.  These binaries were not produced in Run 1 because angular momentum loss during stable mass transfer drove these systems into contact, resulting in a merger.  Additionally, there are some sdB + A or F dwarf binaries formed after a CE phase that begins while the primary is in the HG.  These systems began CE evolution while the stars were more widely separated than the same systems in Run 1, a result of the diminished angular momentum loss during RLOF.       

\autoref{fig:han3}b shows the population produced by Run 10, where we have allowed stars to ignite helium shy of the TRGB.  The post-CE population produced as a result is nearly identical to that produced in Run 2 because these systems did not undergo stable mass transfer and are therefore unaffected by the altered angular momentum loss condition.  We have plotted Run 12 in \autoref{fig:han3}c.  In this run we have changed both $q_{\rm crit}$ and the the binding energy.  By comparing this run to Run 4 (\autoref{fig:han1}d), one can see that reducing the angular momentum carried away by ejected material has resulted in a slight decrease in the number of sdB + G dwarf systems with $P > 10$ days.  Without the angular momentum loss, mass transfer ended in these systems before the primary's envelope was completely stripped.  Finally, in \autoref{fig:han3}d, we show Run 13 and the effect of forcing non-conservative mass transfer by setting $f_{\rm RLOF} = 0.5$.  As with Run 6 (\autoref{fig:han1}f), $T_{\rm eff}$ is lower for the companions in this case because they have not accreted as much material and are therefore less massive.  However, since the lost material is removing far less angular momentum from the system in Run 13 than in Run 6, the orbital periods of the binaries produced by stable RLOF mass transfer in Run 13 are only slightly reduced, with most remaining longer than 10 days.              

Comparing \autoref{fig:ic3} ($\gamma = -1$) with \autoref{fig:ic1} ($\gamma = 1$) reveals that the amount of specific angular momentum carried away by ejected material only weakly impacts which initial binaries will become sdB binaries.  While these runs were useful to illustrate how changing $\gamma$ affects the population of sdB + MS binaries produced in our BPS models, \autoref{fig:wd3} shows that many of these runs do not reproduce the observed systems.        

\subsection{Distributions}
\label{sec:dists}
Another useful way to investigate the BPS populations is to look at the distributions of their properties.  \autoref{fig:teff_hist} shows the distribution of the companion's $\log~T_{\rm eff}$ in each of the runs discussed above.  These distributions are not readily apparent in Figures 3, 6, and 9 due to severe overplotting in some regions. The bin-size is 0.1 in $\log~T_{\rm eff}$, with bin edges centered at multiples of 0.1 in $\log~T_{\rm eff}$. There are two distinct groups, one at high $\log~T_{\rm eff}$ corresponding to the HG RLOF systems and one at low $\log T_{\rm eff}$ due to the RGB CE II systems.  The former group is always present.  The RGB CE II group is dominant when present, but only appears in runs with $f_{\rm He} = 0.95$.  The region between these groups can be populated by systems formed through either the RGB CE I channel or the RGB RLOF channel.  When the RGB CE II group is not present, the post-RGB systems dominate the distribution for $T_{\rm eff} < 3.85$.  The same grouping, by formation channel, is seen in the age distributions of these runs, which are shown in \autoref{fig:age_hist}.  Most of the post-RGB binaries are formed in binaries with $M_{10} < 2 M_{\sun}$ and they produce the distinct feature at high age seen in Runs 1, 17, and 9.            

The masses of the sdBs in the modeled populations are shown in \autoref{fig:msdb_hist}.  When $f_{\rm He} = 0.95$, the distribution peaks near the canonical value of $0.48~ M_{\sun}$.  The majority of sdBs in these runs are formed through the RGB CE II channel by progenitors that ignite He degenerately near the TRGB.  Stars that ignite He non-degenerately on the RGB produce the sdBs that populate both the low and high mass tails of these distributions.  Primaries with $M_{10} \sim 2~M_{\sun}$ that begin transferring mass in the HG can become low mass sdBs with $0.32 \la M_{\rm sdB} \la 0.4$.  Interestingly, our models still produce high mass sdBs, despite the fact that we do not follow mergers.  More massive ($M_{10} \ga 4~M_{\sun}$) primaries undergoing stable RLOF mass transfer that begins in the HG yield sdBs with masses as high as $0.79~M_{\sun}$.  In each run, the post-RGB stars populate a separate, low-mass region.     

\section{Constraints on BPS Parameters}
\label{sec:constraints}
In \autoref{section:gen} we defined the run suitability parameter for each run, $\rho$, that is the fraction of observed short period sdB binaries spanned by the short period sdB + WD and/or sdB + M dwarf binaries in each BPS model.  Twelve of the parameters sets explored here were able to reproduce more than 90\% of systems in the observed sample, despite making very different predictions about the sdB + MS population.  For any value of any one parameter, a combination of the others exists which is qualitatively consistent with the observed sample.  In fact, the only parameter varied in our study that was well constrained by current observations is $f_{\rm He}$.  Only models with $f_{\rm He} = 0.95$ produced sdB + WD and sdB + M dwarf systems with $P > 1$ day, which are required by observations.  However, this is not necessarily a statement about the minimum core mass at which helium ignition occurs.  With $f_{\rm He} = 1.0$ a primary that ignites helium degenerately (i.e., one with $M_{1} \la 2~M_{\sun}$) could only form an sdB if it filled its Roche lobe at exactly the same time it reached the TRGB and ignited helium.  This scenario depends on many assumptions built into \bse~about single star evolution, the size of the Roche lobe, and the onset and duration of unstable mass transfer, such that there is a narrow range of $q_{0}$ and $P_{0}$ with a given $M_{10}$ for which it is possible, as was pointed out by \citet{Mengel:1976}. (Compare panels a and b in Figures 4 and 5 for example).  In practical terms, changing $f_{\rm He}$ extends the range of initial orbital periods for which a system with a primary of $M < 2~M_{\sun}$ and $q > q_{\rm crit}$ will produce a binary containing an sdB, thus adding some flexibility to the \bse~algorithms.  This flexibility is essential to reproducing the observed population, but it is impossible to determine exactly which physical process or combination of processes requires this freedom. Further observations are needed to better constrain binary evolution parameters and go beyond the limits of existing studies.                                   

Still, twelve combinations of the remaining parameters produced a population consistent with the observed sample.  Models that used the  LVK11 binding energies required values of \ace~$> 0.75$ to produce binaries similar the observed systems with the longest periods and highest companion masses.  Nevertheless, fixing the value of any one parameter, we can find a set of values for the remaining parameters that produces a population consistent with the observed sample. Note, however, that if the BPS parameters are correlated, this may not be possible.  We conclude that the short period sdB + WD and sdB + M dwarf systems that constitute the presently well-explored sample are fairly insensitive to the parameterizations used in our BPS models, and propose that observations of sdB + early F dwarf binaries could better them.                  

\subsection{The Advantage of sdB + Early F dwarf Binaries}
Our models suggest that the sdB + early F dwarf ($7350 > T_{\rm eff} > 6700$) binary population can discriminate between different binary evolution scenarios.  Stars in this temperature range are cool enough that the sdB will produce a measurable far-ultraviolet excess, without being so cool as to make selection based on optical/IR colors
difficult.  We show normalized histograms of the {\em helium burning} sdB + early F binary period distributions for the twelve runs with $\rho > 0.9$ in \autoref{fig:fhist} (hatched histograms).  To ensure that our models properly sampled the period distribution of sdB + early F dwarf binaries, we ran additional, ``focused'' BPS models for each of these twelve runs.  We drew $10^{6}$ initial binaries from the same distributions described above, but we restricted the range of $M_{10}$, $a$, and $q_{0}$ to maximize the number of sdB + early F dwarf binaries.  These runs produced 200--1100 times as many sdB + early F dwarf binaries as those listed in \autoref{table}. We find good agreement between period distributions in these models and those described above and we plot the distributions for the ``focused'' models.      

These twelve sets of parameters predict populations of sdB + early F dwarf binaries that are useful in constraining $q_{\rm crit}$, $\gamma$, and \ace.  First, these systems will only have $P > 75$ days if they are formed through stable RLOF mass transfer that begins while the primary is on the RGB, see \autoref{fig:fhist} panels b, c, d, g, h, k, and l.  This scenario is only possible if $q_{\rm crit} > 1$, which is almost never the case given \autoref{eqn:qcrit}.  The presence of sdB + F dwarf binaries with $P > 75$ days would be evidence that mass transfer beginning on the RGB can be stable.  Furthermore, if the mass ejected from the binary during non-conservative mass transfer only carries away the specific angular momentum of the primary ($\gamma = -1)$, the orbital periods of the sdB + early F binaries would be even longer, $P > 250$ day, see \autoref{fig:fhist} panels d, k, and l. 

On the other hand, the absence of long period ($P > 75$ day) sdB + F dwarf binaries would indicate that mass transfer beginning on the RGB is unstable and that such systems cannot avoid CE evolution.  If systems that begin mass transfer while the primary is on the RGB do undergo CE evolution, they will only survive as short period ($P < 5$ days) sdB + F dwarf binaries if \ace $> 0.75$, see \autoref{fig:fhist} panels c--l.  From \autoref{fig:fhist}a it appears that there are short period sdB + F dwarf binaries that survive with \ace~= 0.75, however the figure is misleading because there is only a single sdB + early F dwarf binary produced in Run 2 (see \autoref{fig:han1}b).  

In panel j of \autoref{fig:fhist}, we see a population of sdB + F binaries with $P > 10$ days.  These binaries form through stable RLOF mass transfer that begins while the primary is in the HG, but only if material ejected from the system carries away the specific angular momentum of the primary (see panels a and b of \autoref{fig:han3}).  These binaries should be distinguishable from those formed through stable mass transfer beginning on the RGB because they all have $P < 70$ day.  The constraints placed on binary evolution scenarios  by each possible population of {\em helium burning}  sdB + F early dwarf binaries  can be summarized as follows:
\begin{enumerate}
\item{No sdB + early F binaries: unstable mass transfer on the RGB and \ace~$\la0.75$}
\item{Only sdB + early F binaries with $P \ga 75$ days: stable mass transfer on the RGB, \ace~$\la 0.75$}
\begin{itemize}
\item[a)] If sdB + early F binaries have $P \ga 250$ day then $\gamma = -1$, otherwise $\gamma =1$
\end{itemize}
\item{Only sdB + early F binaries with $P \la 75$: unstable mass transfer on RGB and \ace~$\ga 0.75$}
\begin{itemize}
\item[a)] If sdB + early F binaries have $P \ga 10$ day then $\gamma = -1$, otherwise $\gamma =1$
\end{itemize}
\item{Long and short period sdB + F: stable mass transfer on the RGB and \ace~$\la 0.75$}
\begin{itemize}
\item[a)] If sdB + early F binaries have $P \ga 250$ day then $\gamma = -1$, otherwise $\gamma =1$
\end{itemize}
\end{enumerate}         
Clearly, the {\em helium burning} sdB + early F dwarf binary population has exceptional diagnostic power.  However, such binaries are largely absent from catalogs because sdBs are discovered in optical color surveys that search for faint blue stars;  sdB + early F binaries may be missed because the luminous companion outshines the sdB \citep{Wade:2010}.          

There is an additional complication to studying the sdB + F binaries. Many of the post-RGB objects have F type companions (see Figures 3, 6, and 9), and we plot their period distribution (rescaled to be consistent with the ``focused'' BPS models) as the red histogram in \autoref{fig:fhist}.  This overlapping population requires that the nature of the F star's companion be carefully confirmed before the system can be used to constrain binary evolution models, because the period distribution of post-RGB binaries is not as sensitive to the parameters explored here.        

\section{Comparison with Previous Studies}
\label{sec:comp}
In Figures 3, 6, and 9 we have indicated the region in $\log P- \log T_{\rm eff}$ space occupied by the binaries predicted in the best fit model chosen in H03 (see their Figure 15).  In our Run 6, we used parameters similar to H03 and were able to reproduce many features of the binary population presented there.  The orbital periods predicted in our model range from 1 hr to 663 days which is similar to the 0.5 hr to 500 day range predicted in H03.   Although the calculations of \citet{Han:2002} show that it may be possible to create sdB + MS binaries with $P > 1000$ days, neither our BPS models nor the H03 BPS models produce such systems.  One notable difference is that our model predicted systematically lower temperatures for the companions in systems that formed through stable RLOF mass transfer.  This difference can be accounted for, in part, by the different approaches used in the BPS calculations.  H03 assumed that RLOF mass transfer was instantaneous, and applied the results of detailed binary calculations presented in \citet{Han:2002} to determine the minimum core mass required at the onset of RLOF to produce an sdB.  If RLOF began when the primary's core was massive enough ($0.28 M_{\sun} \le m_{\rm c} \le 0.39 M_{\sun}$, depending on the star's total mass), the entire envelope was stripped and half of the envelope mass was transferred to the companion (Z. Han, private communication).  Because the core was not allowed to grow during RLOF, this resulted in less massive sdBs and excess material being transferred to the companions.  Our \bse~models allowed the mass donor to evolve during RLOF which decreases the mass of the envelope and reduces the amount of material transferred to the companion by $\sim 0.1~M_{\sun}$.  The lower temperatures predicted in our models result in a smaller ``gap'' in the temperature distribution for systems with $P < 10$ days.  In H03 there are no systems with $3.8 < \log T_{\rm eff} < 4.0$ at periods shorter than 10 days.  In Run 6, the gap only extends from $\log T_{\rm eff} = 3.8 - 3.85$.  

The nature of the post-CE systems in some of our models differed greatly from the predictions of H03 and \citet{Nelemans:2010}.  In the models that take $q_{\rm crit}$ from \autoref{eqn:qcrit}, the periods of the post-CE binaries extend to much higher values than those predicted by H03.  The maximum period of these post-CE systems ranges from 181 days with \ace~= 0.75 to 543 days with \ace~= 3.0 (e.g., Run 34).  These long period post-CE systems are produced by binaries with wide initial separations and G star primaries.  The BPS models presented in \citet{Nelemans:2010} use a different prescription for the CE ejection process in which it is possible for the period of a system to increase during during the CE phase.  This results in post-CE binaries with $P > 1000$ days, far longer than the long period systems predicted in our models.  The short period, post-CE systems with G or later companions predicted by our models are consistent with those predicted by H03.  This is contradicted by the lack of observed short period sdB + G dwarf systems.  However, some of our models show there are far more sdB + G dwarf systems with long periods than with short, which is consistent with the findings of \citet{Green:2001} and \citet{Copperwheat:2011}.      

\section{Conclusion}  
We have presented a set of BPS models that explored the population of sdBs in binaries with hydrogen burning companions.  We found that this population is highly sensitive to the parameters used to simulate binary evolution.  Plausible variations in the assumptions about the minimum core mass for helium ignition, the envelope binding energy, the common envelope ejection efficiency, the amount of mass and angular momentum lost during stable mass transfer, and the criteria for stable mass transfer result in populations of sdB + MS binaries with drastically different orbital period and companion temperature distributions. Our \bse~models also suggest that the population of sdB + WD and sdB + M dwarf binaries used to constrain the values of these parameters cannot do so unambiguously.  Twelve very distinct parameter sets can reproduce over 90\%
of the systems in the currently well-studied sample.

The period distribution of {\em helium burning} sdB + early F dwarf binaries, on the other hand, can differentiate between the binary evolution scenarios described by these parameterizations.  The presence of sdB + early F dwarf binaries with $P < 5$ days would indicate that orbital energy is efficiently transferred to the CE.  Additionally, the existence of sdB + early F dwarf binaries with $P > 75$ day would require that mass transfer beginning while the primary is on the RGB is stable.  Details of the sdB + early F dwarf binary distribution also probe angular momentum loss during non-conservative mass transfer.  Additionally, the models predict a population of binaries containing post-RGB objects that lie in the sdB ``box'' that is, like the population of sdB + early F dwarf binaries, largely absent from catalogs.                    

\acknowledgments

We thank Zhanwen Han for helpful discussions.  This material is based upon work supported in part by the National Science Foundation under Grant No. AST-0908642. DC gratefully acknowledges support from the National Aeronautics Space Administration (NASA) through Chandra Award Number TM8-9007X issued by the Chandra X-ray Observatory Center, which is operated by the Smithsonian Astrophysical Observatory for and on behalf of NASA under contract NAS8-03060 and the PSU Academic Computing Fellowship. RK gratefully acknowledges funding from NASA Astrobiology Institute's Virtual Planetary Laboratory lead team, supported by NASA under cooperative agreement NNH05ZDA001C, and the Penn State Astrobiology Research Center. ROS is supported by NSF award PHY-0970074, the Bradley Program Fellowship, and the UWM Research Growth Initiative.

\clearpage
\begin{landscape}
\begin{deluxetable}{ccrcccccrrrrrrrrrrc}
\tablecolumns{19}
\tablewidth{0pt}
\tablecaption{\texttt{BSE} Runs \label{table}}
\tablehead{
\colhead{}& \colhead{}& \colhead{}& \colhead{}& \colhead{$q_{\rm crit}$}&\colhead{$q_{\rm crit}$}& \colhead{}&\colhead{}&\colhead{unique}&\colhead{resampled}&\colhead{RLOF}& \colhead{RLOF}& \colhead{CE}&\colhead{CE} & \colhead{} &\colhead{} &\colhead{} &\colhead{post} &\colhead{}\\
\colhead{Run}	&	\colhead{\ace} &\colhead{ $\gamma$} & \colhead{\frlof}& \colhead{RGB} &\colhead{HG}& \colhead{BE}&\colhead{$f_{\rm He}$} &\colhead{systems} & \colhead{systems} & \colhead{HG} & \colhead{RGB} & \colhead{HG} & \colhead{RGB} & \colhead{other} &\colhead{MS\tablenotemark{a}}&\colhead{WD}&\colhead{RGB} &\colhead{$\rho$}\\
\colhead{(1)}&\colhead{(2)}&\colhead{(3)}&\colhead{(4)}&\colhead{(5)}&\colhead{(6)}&\colhead{(7)}&\colhead{(8)}&\colhead{(9)}&\colhead{(10)}&\colhead{(11)}&\colhead{(12)}&\colhead{(13)}&\colhead{(14)}&\colhead{(15)}&\colhead{(16)}&\colhead{(17)}&\colhead{(18)}&\colhead{(19)}}
\startdata
1 & 0.75 & 1 & \bse \tablenotemark{b} & \bse \tablenotemark{c} & 4.0 & \bse & 1.00 & 711 & 4752  	& 65.85 & 0.00 & 0.00 & 0.55 & 17.59& 83.99	& 4.95 & 11.06 & 0.05\\
2 & 0.75 & 1 & \bse & \bse & 4.0 & \bse & 0.95 & 9334 & 182649 & 1.71 & 0.00 & 0.00 & 96.68 & 0.46& 98.86  & 0.86 & 0.28 & 0.98\\
3 & 0.75 & 1 & \bse & \bse & 4.0 & LVK10 & 0.95 & 9163 & 173620  	& 1.80 & 0.00 & 0.00 & 96.92 & 0.48	& 99.21	& 0.48 & 0.31 & 0.85\\
4 & 0.75 & 1 & \bse & 1.5 & 3.2 & LVK10 & 0.95 & 9878 & 182190  	& 2.20 & 6.36 & 0.00 & 88.33 	& 0.46& 97.35 & 2.28 & 0.37 & 0.72\\
5 & 0.75 & 1 & 0.5 & 1.5 & 3.2 & LVK10 & 0.95 & 10152 & 181097 & 2.15 & 6.29 & 0.00 & 88.86& 0.08& 97.38 	& 1.75 & 0.86 & 0.77\\
\bf 6\tablenotemark{d} & \bf 0.75 & \bf 1 & \bf 0.5 & \bf 1.5 & \bf 3.2 & \bse & \bf 0.95 & \bf 10407 & \bf 191823  & \bf 2.03 & \bf 5.94 & \bf 0.00 & \bf 88.12 &\bf 0.07& \bf 96.17 	& \bf 3.02 & \bf 0.81 & \bf 0.91\\
7 & 0.75 & 1 & 0.5 & 1.5 & 3.2 & \bse & 1.00 & 1740 & 12536 	& 31.11 & 36.13 & 0.00 & 0.21 & 1.12 & 68.56 	& 19.02 & 12.42 & 0.08\\
8 & 0.75 & 1 & 0.5 & 1.5 & 3.2 & LVK10 & 1.00 & 1676 & 10482 & 37.21 & 43.21 & 0.00 & 0.00 & 1.34& 81.75 		& 3.31 & 14.94 & 0.06\\
9 & 0.75 & -1 & \bse & \bse & 4.0 & \bse & 1.00 & 2461 & 14962 & 65.25 & 0.00 & 0.01 & 0.17 & 6.34	& 72.84 		& 1.71 & 25.45 & 0.08\\
10 & 0.75 & -1 & \bse & \bse & 4.0 & \bse & 0.95 & 11081 & 192704	& 5.07 & 0.00 & 0.00 & 91.64 & 0.49& 97.28 	 	& 0.74 & 1.98 & 0.77\\
11 & 0.75 & -1 & \bse & \bse & 4.0 & LVK10 & 0.95 & 10891 & 183643 & 5.32 & 0.00 & 0.00 & 91.63 & 0.52 & 97.47	& 0.45 & 2.09 & 0.78\\
12 & 0.75 & -1 & \bse & 1.5 & 3.2 & LVK10 & 0.95 & 10779 & 187803& 4.21 & 6.01 & 0.00 & 85.69 & 0.51 & 96.41 	& 2.46 & 1.13 & 0.71\\
13 & 0.75 & -1 & 0.5 & 1.5 & 3.2 & LVK10 & 0.95 & 11801 & 194591  & 4.48 & 6.11 & 0.00 & 82.70 & 0.91& 94.19	& 3.22 & 2.59 & 0.74\\
14 & 0.75 & -1 & \bse & 1.5 & 3.2 & \bse & 0.95 & 12036 & 203800 	& 4.28 & 5.83 & 0.00 & 82.94 & 0.87 & 93.91 	& 3.61 & 2.48 & 0.89\\
15 & 0.75 & -1 & 0.5 & 1.5 & 3.2 & \bse & 1.00 & 3173 & 21137 & 41.23 & 23.12 & 0.00 & 0.12 & 8.35 & 72.82 	& 3.31 & 23.87 & 0.05\\
16 & 0.75 & -1 & 0.5 & 1.5 & 3.2 & LVK10 & 1.00 & 3135 & 20533& 42.44 & 23.80 & 0.00 & 0.00 & 8.59  & 74.84 	& 0.59 & 24.57 & 0.00\\
17 & 1.50 & 1 & \bse & \bse & 4.0 & \bse & 1.00 & 1986 & 16740 & 18.69 & 0.00 & 0.01 & 55.43 & 5.14 & 79.46 		& 17.67 & 2.87 & 0.25\\
18 & 1.50 & 1 & \bse & \bse & 4.0 & \bse & 0.95 & 10732 & 205101& 1.53 & 0.00 & 0.00 & 96.10 & 0.42& 98.06  	& 1.70 & 0.23 & 0.75\\
19 & 1.50 & 1 & \bse & \bse & 4.0 & LVK10 & 0.95 & 9479 & 187221& 1.67 & 0.00 & 0.00 & 96.13 & 0.45& 98.25  	& 1.47 & 0.28 & 0.88\\
20 & 1.50 & 1 & \bse & 1.5 & 3.2 & LVK10 & 0.95 & 10485 & 198744 & 2.02 & 5.83 & 0.00 & 86.67 & 0.42& 94.94 	& 4.72 & 0.34 & 0.78\\
21 & 1.50 & 1 & 0.5 & 1.5 & 3.2 & LVK10 & 0.95 & 10491 & 197628	& 1.97 & 5.77 & 0.00 & 87.16 & 0.07 & 94.97  	& 4.24 & 0.79 & 0.83\\
22 & 1.50 & 1 & 0.5 & 1.5 & 3.2 & \bse & 0.95 & 11549 & 218153& 1.79 & 5.23 & 0.00 & 86.38 & 0.08& 93.47  	& 5.83 & 0.70 & 0.91\\
23 & 1.50 & 1 & 0.5 & 1.5 & 3.2 & \bse & 1.00 & 2730 & 28199 	& 13.86 & 16.06 & 0.00 & 30.13 & 0.59& 60.63 	& 33.95 & 5.42 & 0.37\\
24 & 1.50 & 1 & 0.5 & 1.5 & 3.2 & LVK10 & 1.00 & 1776 & 15474 & 25.20 & 29.27 & 0.00 & 0.05 & 0.90& 55.42 	& 34.45 & 10.13 & 0.12\\
25 & 1.50 & -1 & \bse & \bse & 4.0 & \bse & 1.00 & 3792 & 28023 & 34.84 & 0.00 & 0.00 & 33.11 & 3.48& 75.02  	& 11.54 & 13.44 & 0.28\\
26 & 1.50 & -1 & \bse & \bse & 4.0 & \bse & 0.95 & 12538 & 216384	& 4.51 & 0.00 & 0.00 & 91.09 & 0.45 & 96.52 	& 1.74 & 1.74 & 0.78\\
27 & 1.50 & -1 & \bse & \bse & 4.0 & LVK10 & 0.95 & 11227 & 197282 & 4.95 & 0.00 & 0.00 & 91.23 & 0.48& 96.66 	& 1.41 & 1.93 & 0.88\\
28 & 1.50 & -1 & \bse & 1.5 & 3.2 & LVK10 & 0.95 & 11401 & 204381 & 3.87 & 5.53 & 0.00 & 84.28 & 0.46& 94.13  	& 4.83 & 1.03 & 0.78\\
29 & 1.50 & -1 & 0.5 & 1.5 & 3.2 & LVK10 & 0.95 & 12115 & 208477	& 4.18 & 5.70 & 0.00 & 82.62 & 0.85 & 93.35 	& 4.24 & 2.42 & 0.85\\
30 & 1.50 & -1 & \bse & 1.5 & 3.2 & \bse & 0.95 & 13220 & 228223	& 3.82 & 5.21 & 0.00 & 82.56 & 0.78  & 92.38 		& 5.43 & 2.20 & 0.91\\
31 & 1.50 & -1 & 0.5 & 1.5 & 3.2 & \bse & 1.00 & 4207 & 34600 & 25.19 & 14.12 & 0.00 & 24.55 & 5.17 & 69.03 	& 16.49 & 14.48 & 0.26\\
32 & 1.50 & -1 & 0.5 & 1.5 & 3.2 & LVK10 & 1.00 & 3204 & 22803 & 38.22 & 21.43 & 0.00 & 0.03 & 7.74& 67.42 	& 10.50 & 22.08 & 0.08\\
33 & 3.00 & 1 & \bse & \bse & 4.0 & \bse & 1.00 & 8169 & 236524 & 1.34 & 0.00 & 0.00 & 92.31 & 0.39& 94.35 	& 5.46 & 0.19 & 0.74\\
34 & 3.00 & 1 & \bse & \bse & 4.0 & \bse & 0.95 & 16979 & 430671	& 0.73 & 0.00 & 0.00 & 95.39 & 0.22& 96.52 		& 3.38 & 0.10 & 0.98\\
35 & 3.00 & 1 & \bse & \bse & 4.0 & LVK10 & 0.95 & 11161 & 234647 & 1.33 & 0.00 & 0.00 & 95.85 & 0.37& 97.56		& 2.23 & 0.21 & 0.92\\
36 & 3.00 & 1 & \bse & 1.5 & 3.2 & LVK10 & 0.95 & 12128 & 260984& 1.54 & 4.44 & 0.00 & 82.43 & 0.33& 88.74 	& 11.01 & 0.25 & 0.89\\
37 & 3.00 & 1 & 0.5 & 1.5 & 3.2 & LVK10 & 0.95 & 11913 & 246390 	& 1.58 & 4.63 & 0.00 & 87.31 & 0.07& 93.59 		& 5.79 & 0.62 & 0.91\\
38 & 3.00 & 1 & 0.5 & 1.5 & 3.2 & \bse & 0.95 & 17611 & 433900 & 0.91 & 2.63 & 2.63 & 92.00 & 0.05& 95.65  	& 4.00 & 0.34 & 0.98\\
39 & 3.00 & 1 & 0.5 & 1.5 & 3.2 & \bse & 1.00 & 8753 & 239108  & 1.65 & 1.89 & 1.89 & 89.75 & 0.10& 93.50 	& 5.88 & 0.62 & 0.75\\
40 & 3.00 & 1 & 0.5 & 1.5 & 3.2 & LVK10 & 1.00 & 3083 & 53220& 7.33 & 8.51 & 0.00 & 59.40 & 0.33& 75.57	 	& 21.57 & 2.87 & 0.46\\
41 & 3.00 & -1 & \bse & \bse & 4.0 & \bse & 1.00 & 10007 & 251117	& 3.89 & 0.00 & 0.00 & 86.94 & 0.42& 92.44 		& 6.08 & 1.48 & 0.77\\ 
42 & 3.00 & -1 & \bse & \bse & 4.0 & \bse & 0.95 & 18818 & 445368 & 2.19 & 0.00 & 0.00 & 92.24 & 0.23 & 95.34 	& 3.82 & 0.84 & 1.00\\
43 & 3.00 & -1 & \bse & \bse & 4.0 & LVK10 & 0.95 & 12932 & 245133& 3.98 & 0.00 & 0.00 & 91.75 & 0.40& 96.14 	& 2.33 & 1.54 & 0.92\\
44 & 3.00 & -1 & \bse & 1.5 & 3.2 & LVK10 & 0.95 & 13066 & 267069 & 2.96 & 4.23 & 0.00 & 80.55 & 0.37& 88.11  	& 11.11 & 0.78 & 0.88\\
45 & 3.00 & -1 & 0.5 & 1.5 & 3.2 & LVK10 & 0.95 & 13607 & 256735 & 3.39 & 4.63 & 0.00 & 83.79 & 0.70& 92.52  	& 5.53 & 1.95 & 0.91\\
46 & 3.00 & -1 & \bse & 1.5 & 3.2 & \bse & 0.95 & 19373 & 445589 	& 1.96 & 2.67 & 2.67 & 89.59 & 0.42& 94.73 		& 4.15 & 1.11 & 0.98\\
47 & 3.00 & -1 & 0.5 & 1.5 & 3.2 & \bse & 1.00 & 10321 & 246973 	& 3.53 & 1.98 & 1.98 & 86.89 & 0.75 & 93.33 	& 4.66 & 2.01 & 0.72\\
48 & 3.00 & -1 & 0.5 & 1.5 & 3.2 & LVK10 & 1.00 & 4581 & 59674 	& 14.60 & 8.19 & 0.00 & 52.97 & 3.01 & 78.78 	& 12.82 & 8.39 & 0.43\\
\enddata
\tablenotetext{1}{ Columns (11) - (18) are percentages of the number of systems given in column (10) formed by each channel.  The values in columns (11)-(15) sum to the value in column (16).}
\tablenotetext{2}{See \autoref{eqn:frlof}}
\tablenotetext{3}{See \autoref{eqn:qcrit}}
\tablenotetext{4}{This run most closely resembles the best fit run in H03.}
\end{deluxetable}
\clearpage
\end{landscape}
\end{document}